\documentclass[journal,final,a4paper,twocolumn,10]{IEEEtran}

\ifCLASSINFOpdf
\else
\fi

\usepackage{amsmath,graphicx}
\usepackage{amssymb,epsfig,color}

\newcommand{\vm}[1]{\ensuremath{\bm{#1}}}

\renewcommand{\exp}[1]{\ensuremath{\text{e}^{#1}}}
\renewcommand{\log}[1]{\ensuremath{\text{log}\!\left(#1\right)}}

\usepackage{bm}
\usepackage[utf8]{inputenc}
\usepackage{url}
\usepackage[table]{xcolor}
\usepackage{multirow}
\usepackage{makecell}
\usepackage{tikzsymbols}
\usepackage{algorithm,algpseudocode}
\usepackage{dsfont}

\usepackage{caption}
\usepackage{subfig}
\usepackage{fixltx2e}      
\usepackage{array}

\begin{document}
%

\title{Blind Speech Separation and Dereverberation using Neural Beamforming}

%
%

\author{Lukas~Pfeifenberger and Franz Pernkopf,~\IEEEmembership{Senior Member,~IEEE}
\thanks{Lukas~Pfeifenberger and Franz Pernkopf are with the Intelligent Systems Group at the Signal Processing and Speech Communication Laboratory, Graz University of Technology, Graz, Austria.}
\thanks{This work was supported by the Austrian Science Fund (FWF) under the project number P27803-N15. Furthermore, we acknowledge NVIDIA for providing GPU computing resources.}}


\maketitle

\begin{abstract}

In this paper, we present the \emph{Blind Speech Separation and Dereverberation} (BSSD) network, which performs simultaneous \emph{speaker separation}, \emph{dereverberation} and \emph{speaker identification} in a single neural network. Speaker separation is guided by a set of predefined spatial cues. Dereverberation is performed by using neural beamforming, and speaker identification is aided by embedding vectors and triplet mining. We introduce a \emph{frequency-domain} model which uses complex-valued neural networks, and a \emph{time-domain} variant which performs beamforming in latent space. Further, we propose a \emph{block-online} mode to process longer audio recordings, as they occur in meeting scenarios. We evaluate our system in terms of \emph{Scale Independent Signal to Distortion Ratio} (SI-SDR), \emph{Word Error Rate} (WER) and \emph{Equal Error Rate} (EER).

\end{abstract}
\begin{IEEEkeywords}
Multi-channel speech separation, beamforming, dereverberation, speaker identification, triplet mining
\end{IEEEkeywords}

\IEEEpeerreviewmaketitle

\section{Introduction}
\label{sec:introduction}

Speaker separation and speech enhancement is of paramount significance in many voice applications, such as hands-free teleconferencing or meeting scenarios. Especially in human-machine interfaces, where high-performance Automatic Speech Recognition (ASR) systems are essential, both speech intelligibility and quality play an important role. Fueled by the success of deep learning, both speaker separation and speech enhancement have made major advances over the last years \cite{Wang:may2018}.

When multiple microphones are available, spatial information can be exploited as speaker sources are directional. Mask-based beamforming has been shown to be advantageous for this task \cite{chime3overview, chime4_proceedings}. In particular, a neural network is leveraged to estimate a time-frequency mask of the desired signal \cite{Erdogan:sep16, Heymann:mar16, Pfei:aug17, Pfei:aug2019}. This mask is then used to compute the spatial covariance matrices required to construct a frequency-domain beamformer \cite{speechprocessing1}. This approach has been further extended into the domain of complex numbers, where complex-valued neural networks \cite{Trabelsi:2017} are used to directly estimate complex beamforming weights from noisy observations \cite{Pfei:may2019, Koyama:2019}.

Simultaneously to the success of neural beamforming, single-channel speaker separation techniques have also progressed dramatically. Frequency-domain algorithms such as \emph{Deep Clustering} (DC) \cite{Hershey:2016}, \emph{Permutation Invariant Training} (PIT) \cite{Yu:2017} and \emph{Deep Attractor Network} (DAN) \cite{Chen:2017} rely solely on spectral features. Time-domain algorithms such as \emph{Wave-U-Net} \cite{Stoller:2018}, \emph{TasNet} \cite{Luo:2018} and \emph{Conv-TasNet} \cite{Luo:2019} delivered promising results.

Recently, some of these single-channel algorithms have been combined with a mask-based beamformer. In particular, a neural network estimates a gain mask of the desired signal, which is then used to construct a frequency-domain beamformer, i.e.: \emph{Beam-TasNet} \cite{Ochiai:2020}, \emph{SpeakerBeam} \cite{Delcroix:aug2017, Delcroix:aug2019}, \emph{Neural Speech Separation} \cite{Yoshioka:apr2018}, \emph{Multi-Channel Deep Clustering} \cite{Wang:sep2018, Wang:2018}, and \emph{Convolutional Beamforming} \cite{Nakatani:2020}. More recently, end-to-end multi-channel speech separation has been done entirely in time domain. By leveraging spatial cues among the multi-channel signals such as the Inter-channel Time Difference (ITD) or Inter-channel Phase Difference (IPD), the desired signal is estimated directly in time domain \cite{Chang:2019}. We further extend this approach by addressing the following three issues: 

\subsection{Open number of sources}
Many source separation algorithms are limited to a pre-defined number of sources which they can separate \cite{Yu:2017, Chen:2017, Luo:2018, Luo:2019, Stoller:2018, Ochiai:2020, Yoshioka:2018, Yoshioka:2019}. Exceptions are k-means clustering \cite{Hershey:2016} and \emph{Recurrent Selective Attention Network} (RSAN) \cite{Neumann:2019}. We propose an iterative approach to separate an unknown number of sources, by leveraging the spatial information encoded within the data. As shown in the experiments section, we tested our system with up to four overlapping speakers.

\subsection{Distant speaker separation}
While close-talk speech separation models yield impressive performance, far-field speech separation is still a challenging task \cite{chime5, watanabe2020chime6}. Especially in real-world scenarios, reverberation and echoes cannot be ignored, as they severely degrade speech intelligibility and ASR performance \cite{Yoshioka:2012}. Various deep learning based methods have been proposed for dereverberation \cite{Kinoshita:2017, Williamson:2017, Nakatani:2019}, most of which are based on the Weighted Prediction Error (WPE) algorithm \cite{Yoshioka:2009}. As this algorithm is restricted to the frequency domain, we chose a more general approach which learns to separate overlapping speakers from a reverberated mixture in both time- and frequency-domain.

\subsection{Speaker Identification}
To be useful in real-world applications, a speaker separation algorithm has to be at least \emph{block-online} capable, i.e.; a short block of audio is processed at a time. This implies a permutation problem at block level, requiring \emph{speaker diarization} \cite{Sell:2018, Yu:jul2017}. Therefore, identifying the separated speakers in each block of audio is necessary. A speaker identification algorithm is agnostic to the spoken text, and only relies on the speaker characteristics embedded in the waveform. Embedding vectors are used to map utterances into a feature space where distances correspond to speaker similarity \cite{Snyder:2016}. Typically, i-Vectors \cite{Dehak:2011} or x-Vectors \cite{Snyder:2018} are used for this task. Algorithms such as \emph{Deep Speaker} \cite{Li:2017} rely on contrastive loss or triplet loss to learn embeddings on a very large set of speakers \cite{Schroff:2015, Wang:sep2017, Song:2018, Zhang:2018}. We chose the triplet loss as its performance on small batch sizes is advantageous \cite{Wu:2017}.

In this paper, we introduce our \emph{Blind Speech Separation and Dereverberation} (BSSD) network, which performs \emph{separation}, \emph{dereverberation} and \emph{speaker identification} in a single neural network. The only knowledge required is that of the microphone array geometry. We propose a frequency-domain variant (BSSD-FD), and time-domain variant (BSSD-TD). Our contributions are: Unsupervised speaker localization; separation of each speaker using adaptive beamforming; dereverberation of each source; and speaker diarization using embedding vectors. We evaluate our system in both \emph{offline} and \emph{block-online} mode. Further, we report the performance in terms of SI-SDR, WER and EER against similar state-of-the art algorithms for speaker separation.

\section{System Model}
\label{sec:system_model}

\begin{figure}[!ht]
\centering
\includegraphics[width=0.9\linewidth]{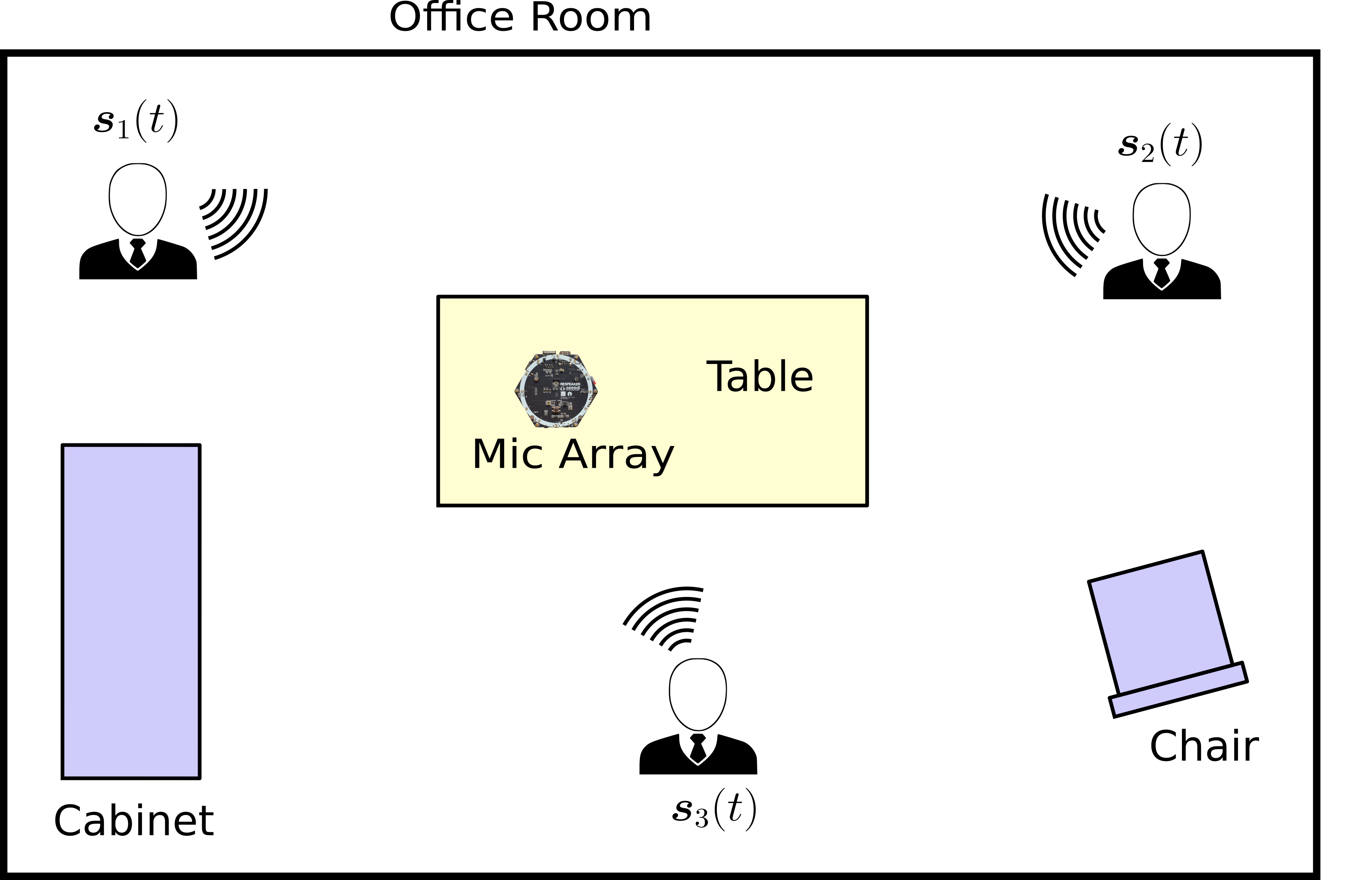}
\caption{Meeting room scenario with three independent speakers.}
\label{fig:room}
\end{figure}

We assume a standard meeting scenario, where multiple speakers may talk simultaneously in an arbitrary room, i.e. an office. The position and number $C$ of the concurrent speakers is unknown. We place a circular microphone array with $M$ microphones in the center of the room, i.e. on a table. Figure \ref{fig:room} provides an example with three speakers. We assume that each speaker has a direct line of sight to the microphone array, i.e. the speaker is not obscured by a corner, or standing in the next room. Each speaker is assumed to be stationary, except for minor movements. Further, the room may have a significant amount of reverberation. Note that we do not assume any diffuse or directional noise sources in this paper. For speech separation in the presence of ambient noise, we refer the interested reader to  \cite{Pfei:aug2019}.

\noindent The signal arriving at the microphone array is composed of an additive mixture of $C$ independent sound sources $\vm{s}_c(t)$. In time domain, the samples of all $M$ microphones at sampling time $t$ can be stacked into a single $M \times 1$ vector, i.e. \

\begin{equation}
  \vm{z}(t) = \sum_{c=1}^{C} \vm{s}_c(t),
\label{eq_mixture}
\end{equation}

\noindent where $\vm{z}(t) = \big[ z(t,1), \ldots, z(t,M) \big]^T$. We use bold symbols for vectors, i.e. $\vm{z}(t)$, and plain symbols for scalars, i.e. $z(t,m)$. The vector $\vm{s}_c(t)$ represents the $c^{th}$ sound source at sample time $t$. Each sound source is composed of a monaural recording $s_c(t)$ convolved with a \emph{Room Impulse Response} (RIR) denoted by $\vm{h}_c(t)$, i.e.\

\begin{equation}
  \vm{s}_c(t) = \vm{h}_c(t) \circledast s_c(t),
  \label{eq_soundsource}
\end{equation}

\noindent where $\circledast$ denotes the convolution operator. The RIR models the acoustic path from a sound source to each microphone as a set of $M$ FIR filters, which includes all reverberations and reflections caused by the room acoustics \cite{roomacoustics1}. Modern office rooms are made  of laminate flooring and concrete walls, which have a low acoustic absorption coefficient. Consequently, the reverberation time $RT_{60}$ may be very large, which significantly affects the performance of speech separation and speech recognition algorithms \cite{chime5, watanabe2020chime6, Yoshioka:2012}. 

\begin{figure}[!ht]
\centering
\includegraphics[width=0.7\linewidth]{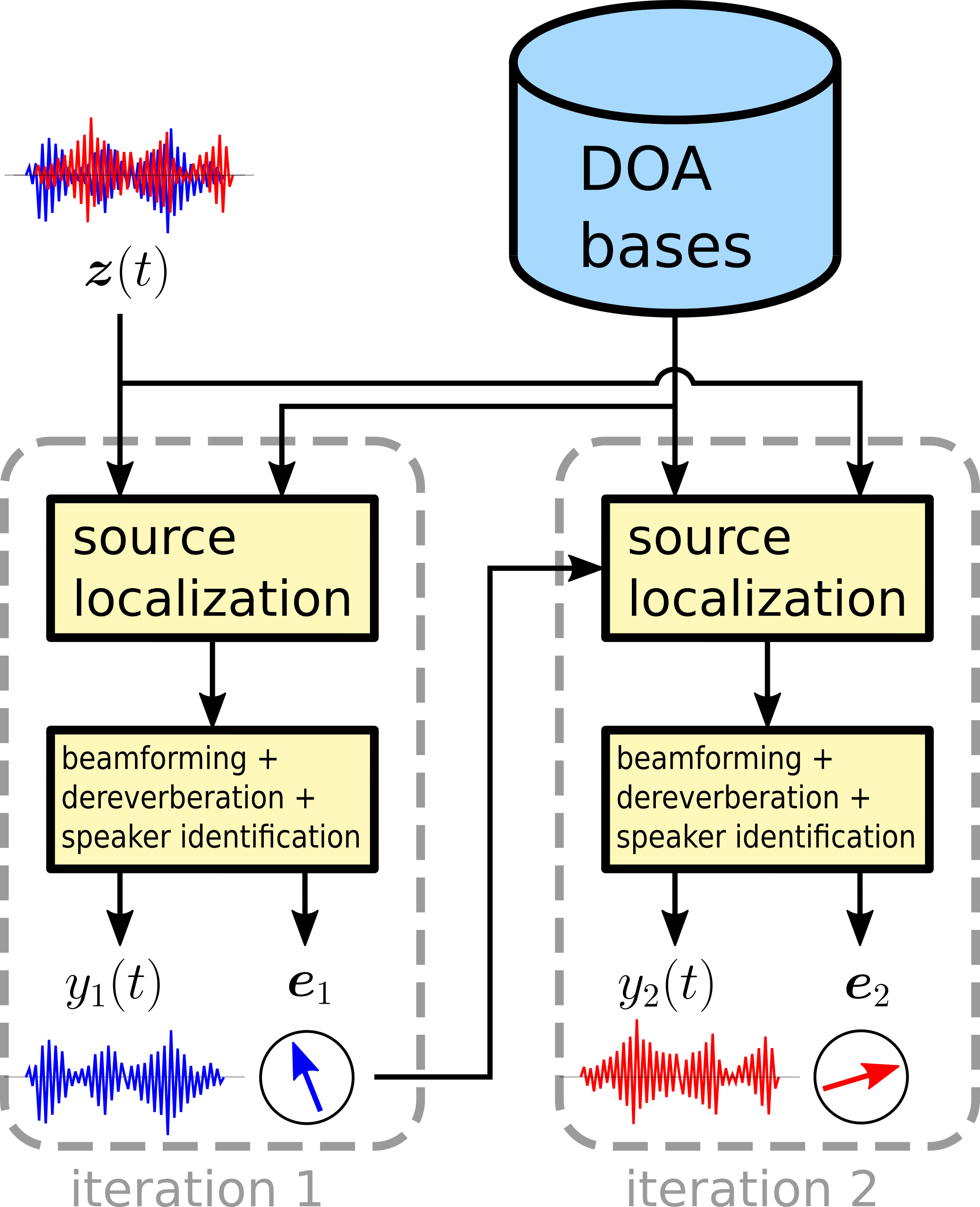}
\caption{Overview of the BSSD system, showing two iterations. During each iteration, the  \emph{localization} module estimates the DOA of a source from a set of pre-defined DOA bases. The DOA is then used to extract and dereverberate the corresponding speech source from the multi-channel input mixture $\vm{z}(t)$. The neural network also assigns a speaker embedding vector $\vm{e}_i$ to each enhanced source $i$.}
\label{fig:system_overview}
\end{figure}

To cope with this environment, we propose the BSSD network, which iteratively extracts an unknown number of speakers from a multi-channel input mixture $\vm{z}(t)$. During each iteration, the Direction Of Arrival (DOA) of the loudest speech source is estimated by a \emph{localization} module, which correlates the input mixture against a pre-defined set of DOA bases. The DOA is subtracted from a spatial speech presence probability map, so that the second-loudest source is extracted during the subsequent iteration. The DOA information is then fed into a Neural Network (NN), which extracts and dereverberates the corresponding speech source. The network also predicts a speaker embedding vector for each extracted speech source, which is used to assign the utterance to a speaker for block-online processing. This iterative process is repeated until no new speaker embedding is found. Figure \ref{fig:system_overview} illustrates two iterations of the BSSD network, which consists of the following modules: \emph{DOA bases}, \emph{localization}, \emph{beamforming and dereverberation}, and \emph{speaker identification}. In the following chapters, we will introduce each module in detail.

\section{DOA Bases}
\label{sec:doa}

As each source in Figure \ref{fig:room} has a direct line of sight towards the microphone array, it is possible to assign a unique DOA to each individual source in the mixture. Even if there is a significant amount of reverberation, there will always be an anechoic component in the RIR (i.e. the earliest peak) that corresponds to the DOA \cite{roomacoustics1}. We therefore define a set of $D$ unique DOA vectors on a unit sphere around the microphone array, where each impinging sound wave is modeled as plane wave, i.e.

\begin{equation}
  V(d,k,m) = \exp{ -i 2 \pi f_k \tau_{d,m} },
\label{eq:doa_vector}
\end{equation}

\noindent where $f_k$ is the frequency for index $k$ and $\tau_{d,m}$ is the time delay from a point on the sphere to the $m$\textsuperscript{th} microphone, i.e.

\begin{equation}
  \tau_{d,m} = \frac{\sqrt{(x_m-x_d)^2 + (y_m-y_d)^2 + (z_m-z_d)^2}}{c}, 
\end{equation}

\noindent where $c$ is the speed of sound. The cartesian coordinates of the $m$\textsuperscript{th} microphone are denoted by $x_m,y_m,z_m$, and $x_d,y_d,z_d$ are the coordinates of the $d$\textsuperscript{th} point on the sphere. We define these points to be equally distributed on the surface of the sphere using a fibonacci spiral \cite{Duvenhage:2013}, i.e.

\begin{equation}
\begin{aligned}
\phi_d &= g \cdot d,\\
\theta_d &= \arcsin{\frac{d}{D-1}},\\
x_d &= \cos{\theta_d} \cos{\phi_d},\\
y_d &= \cos{\theta_d} \sin{\phi_d},\\
z_d &= \sin{\theta_d},\\
\end{aligned}
\label{eq:doa_sphere}
\end{equation}

\noindent where $g = \pi(3-\sqrt{5})$ is known as the golden angle \cite{Duvenhage:2013}, and $d = 1 \ldots D$ is the DOA index. We use a circular microphone array with $M$ channels. Hence, the array is flat and we cannot distinguish between positive and negative $z$ coordinates. It is therefore sufficient to only use half a sphere for the DOA bases. To assign a DOA index $\hat{d}$ to a given RIR $\vm{h}(t)$, we utilize GCC-PHAT \cite{micarray1}, i.e.

\begin{equation}
\hat{d} = \underset{d}{\text{argmax}} \sum_{k=1}^K \frac{| \vm{H}^H(k) \cdot \vm{V}(d,k) |^2 }{ |\vm{H}(k)|_2^2 }, 
\label{eq:GCC-PHAT_doa}
\end{equation}

\noindent where $\vm{H}$ represents the FFT of the RIR $\vm{h}(t)$. Note that the amplitude of the DOA vector $\vm{V}(d,k)$ is defined as $1$ in Eq. (\ref{eq:doa_vector}).

\section{Source Localization}
\label{sec:localization}

To estimate the direction of a speech source relative to the microphone array, we again use GCC-PHAT to obtain a spatial speech presence probability map for the input mixture $\vm{z}(t)$ and the DOAs $\vm{V}(d,k)$. First, we transform the input mixture to the frequency domain using the Short-Time Fourier Transform (STFT), i.e.

\begin{equation}
\vm{z}(t) \rightarrow \vm{Z}(l,k),
\label{eq:STFT}
\end{equation}

\noindent where $\vm{Z}(l,k)$ contains $M$ samples of frequency bin $k$ and STFT frame index $l$. Next, we compute the spatial speech presence probability map $\gamma \in [0 \ldots 1]$ as:

\begin{equation}
\gamma(l,k,d) = \frac{| \vm{Z}^H(l,k) \cdot \vm{V}(d,k) |^2 }{ |\vm{Z}(l,k)|_2^2 }.
\label{eq:GCC-PHAT}
\end{equation}

\subsection{Spatial Whitening}

\begin{figure}[!ht]
\centering

\begin{minipage}{0.49\linewidth}
\includegraphics[width=1.05\linewidth]{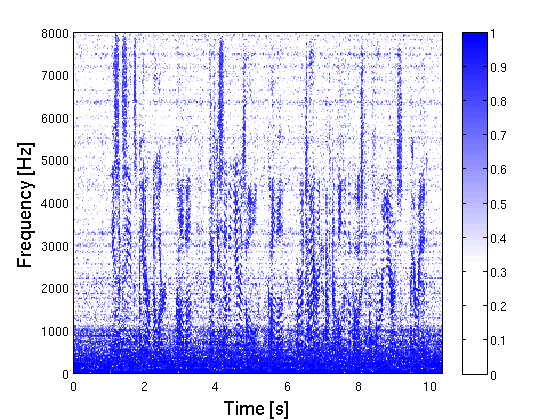}
\end{minipage}
\begin{minipage}{0.49\linewidth}
\includegraphics[width=1.05\linewidth]{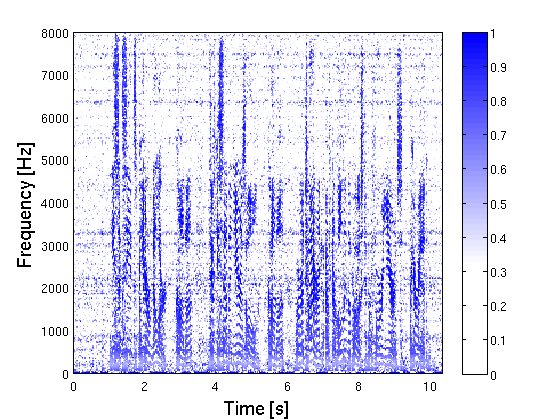}
\end{minipage}

\caption{Effectiveness of spatial whitening at low frequencies. (a) $\gamma(l,k)$ from Eq. (\ref{eq:GCC-PHAT}) for a single speaker. (b) $\gamma_U(l,k)$ from Eq. (\ref{eq:GCC-PHAT-whitened}) with whitening.}
\label{fig:whitening}
\end{figure}

To separate speakers based on their location, Eq. (\ref{eq:GCC-PHAT}) utilizes the IPDs, which are encoded in the phase of the complex-valued input mixture $\vm{Z}(l,k)$. However, it is well known that microphone array recordings are strongly correlated towards low frequencies \cite{micarray1, speechprocessing1, mimo1, micarray2}. This is due to the fact that the wavelength of low frequencies is large compared to the aperture of the microphone array. As a consequence, the IPDs will be small, and the overall separation performance is degraded. To mitigate this effect, we decorrelate the noisy inputs $\vm{Z}(l,k)$ using \emph{Zero-phase Component Analysis} (ZCA) whitening \cite{Bell:1997} from our previous works \cite{Pfei:may2019, Pfei:aug2019}. In particular, we use the whitening matrix 

\begin{equation}
  \vm{U}(k) = \vm{E}_{\Gamma}(k) \vm{D}_{\Gamma}^{-\frac{1}{2}}(k) \vm{E}_{\Gamma}^H(k),
\label{eq_whitening}
\end{equation}

\noindent where $\vm{E}_{\Gamma}$ and $\vm{D}_{\Gamma}$ are $M \times M$ sized eigenvector and eigenvalue matrices of the real-valued spatial coherence matrix of the ideal isotropic sound field $\vm{\Gamma}(k)$ \cite{roomacoustics1}. Its elements are given as $\Gamma_{i,j}(k) = \frac{sin(2 \pi f_k x_{i,j}/c)}{2 \pi f_k x_{i,j}/c}$, and $x_{i,j}$ is the distance between the $i^{th}$ and the $j^{th}$ microphone. To avoid a division by zero, the diagonal elements of $\vm{D}_{\Gamma}$ are loaded with a small constant $\epsilon = 10^{-3}$. We prefer ZCA whitening over PCA whitening, as the ZCA preserves the  orientation of the distribution of the data \cite{Bell:1997}. Using the whitening matrix $\vm{U}(k)$, we rewrite Eq. (\ref{eq:GCC-PHAT}) as

\begin{equation}
\gamma_U(l,k,d) = \frac{| \vm{Z}^H(l,k) \vm{U}^H(k) \cdot \vm{U}(k) \vm{V}(d,k) |^2 }{ | \vm{U}(k) \vm{Z}(l,k) |_2^2 \cdot | \vm{U}(k) \vm{V}(d,k) |_2^2  },
\label{eq:GCC-PHAT-whitened}
\end{equation}

\noindent where $\vm{U}(k) \vm{Z}(l,k)$ can be recognized as the whitened input mixture, and $\vm{U}(k) \vm{V}(d,k)$ as whitened DOA vector. Figure \ref{fig:whitening} demonstrates the effect of spatial whitening. Panel (a) shows $\gamma(l,k)$ for a single speaker and a matching DOA vector. Panel (b) shows $\gamma_U(l,k)$ with whitening. It can be seen that the separation performance is greatly increased for low frequencies. 

\subsection{Speaker Separation and Diarization}

To iteratively estimate the DOA index $\hat{d}$ of all speech sources in the mixture $\vm{Z}(l,k)$, we use the pseudo code in Algorithm \ref{alg:localization}. First, we create a weighted spatial speech presence probability map $\gamma_W(l,k,d)$, using the energy $P_Z(l,k)$ in each time-frequency bin of the input mixture $\vm{Z}(l,k)$. Next, we copy that map into $\gamma_W'(l,k,d)$. Then, we initialize an empty list of speaker embeddings $\mathcal{E}$. During each iteration, we average over the frame and frequency axes of $\gamma_W'(l,k,d)$ to determine its global maximum over the $D$ possible DOAs. The index of the maximum is denoted as $\hat{d}$, which is used as input for the BSSD network, which outputs an estimate of the desired signal $y(t)$ at the direction of the DOA index $\hat{d}$, and a speaker embedding vector $\vm{e}$ for that output. Then, we compare this newly found embedding against the previously stored ones in the list $\mathcal{E}$, using the distance function $\texttt{distance}(\mathcal{E}, \vm{e})$. If the distance is greater than a threshold $\delta$, we append the embedding to the list, and subtract $\gamma_W(l,k,\hat{d})$ from all DOA indices of the weighted spatial speech presence probability map $\gamma_W'(l,k,:)$. This ensures that each speech source is only extracted once \footnote{Note that this algorithm is different to just sorting the DOA indices by energy, as multiple DOA indices may share the energy from the same speaker, due to the limited spatial resolution of the beamformer array.}. If the threshold $\delta$ is not met, the same embedding is already a member of the list $\mathcal{E}$. This may happen due to reflections or sidelobes \cite{micarray1} of the beamformer in the BSSD module. In this case, we stop the iterations and consider all speech sources within the mixture $\vm{z}(t)$ to be extracted. We will discuss the BSSD architecture and the distance function in the following chapters.

\begin{algorithm}[h]
\begin{algorithmic}[1]
  \State $P_Z(l,k) \gets \frac{1}{M} \sum_{m=1}^M |\vm{Z}(l,k,m)|^2$
  \State $\gamma_W(l,k,d) \gets \gamma_U(l,k,d) \cdot P_Z(l,k)$
  \State $\gamma_W'(l,k,d) \gets \gamma_W(l,k,d)$
  \State $\mathcal{E} \gets []$
  \State $\mathcal{Y} \gets []$
  \While{$\textbf{true}$}
    \State $\hat{d} \gets \underset{d}{\texttt{argmax}} \Big( \sum_{l=1}^L \sum_{k=1}^K \gamma_W'(l,k,d) \Big)$
    \State $y(t), \vm{e} \gets \texttt{BSSD}(\vm{z}(t), \hat{d})$
    \State $\mathcal{Y} \texttt{.append}(y(t))$
    \If{$\texttt{distance}(\mathcal{E}, \vm{e}) > \hat{\delta}$}
      \State $\mathcal{E} \texttt{.append}(\vm{e})$
      \State $\gamma_W'(l,k,:) \gets \texttt{max} \Big( \gamma_W'(l,k,:) - \gamma_W(l,k,\hat{d}), 0 \Big)$
    \Else
      \State \textbf{break}
    \EndIf
  \EndWhile
\end{algorithmic}
\caption{Source localization}
\label{alg:localization}
\end{algorithm}

\section{BSSD Network - Frequency Domain}
\label{sec:bssd_fd}

Well-established beamformers such as the \emph{Minimum Variance Distortionless Response} (MVDR) beamformer \cite{Veen:apr88} or the \emph{Generalized Eigenvalue} (GEV) beamformer \cite{Warsitz:Jul07} use the signal statistics (i.e., the power spectral density matrices) to derive a set of beamforming weights $\vm{W}(k) \in \mathbb{C}$ in frequency domain. As those weights are static over time, the signal separation performance is limited especially in reverberant conditions \cite{micarray1}. Therefore, a beamformer is often used in conjunction with a \emph{post-filter} \cite{speechprocessing1}. The post-filter acts as a single-channel gain mask on the output of the beamformer.

In \cite{Pfei:may2019}, we proposed the \emph{Complex-valued Neural Beamformer} (CNBF), which combines the properties of a beamformer and a post-filter using a neural network. Unlike a statistical beamformer, the CNBF estimates a set of individual beamforming weights $\vm{W}(l,k) \in \mathbb{C}$ for each time-frequency bin. Those weights act as a spatio-temporal, complex-valued gain mask, which allows for a higher flexibility in the design of the beamformer, i.e. higher suppression rates or dereverberation. The CNBF uses complex-valued, non-holomorphic activation functions like vector normalization, phase normalization or conjugation. To back-propagate the complex-valued gradient, Wirtinger calculus is used \cite{Amin2011, Bouboulis2011:may2010, Wirtinger2005}. A Tensorflow implementation of the CNBF network can be found at\footnote{https://github.com/rrbluke/CNBF}.

\begin{figure}[!ht]
\centering
\includegraphics[width=0.9\linewidth]{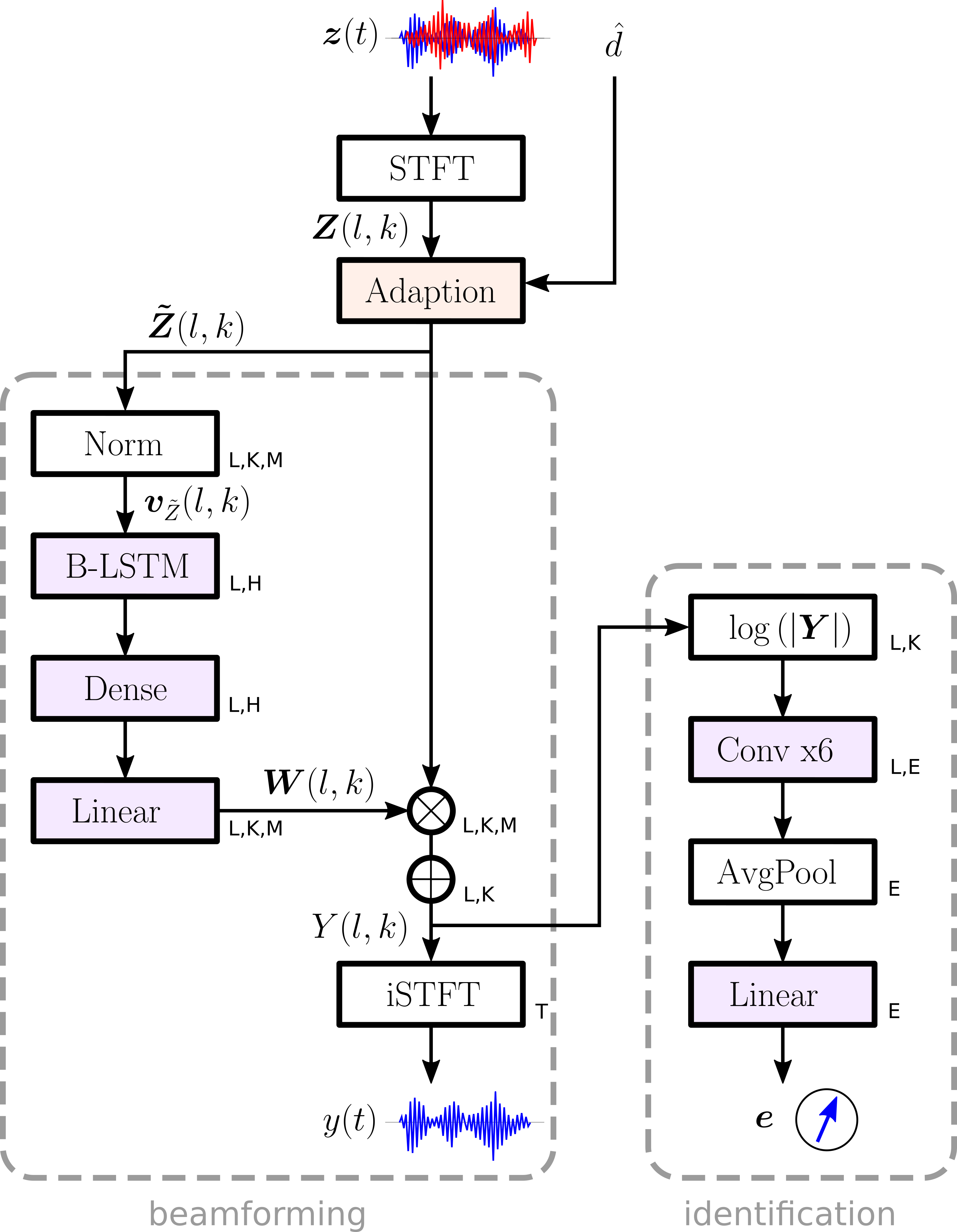}
\caption{Layers of the frequency-domain BSSD-FD network. The left branch performs beamforming and dereverberation, the right branch assigns an embedding vector to the enhanced output signal $y(t)$. The symbols next to each layer denote the dimensionality of the respective output tensor.}
\label{fig:bssd_fd}
\end{figure}

We extend the CNBF to include dereverberation and a speaker embedding vector. Figure \ref{fig:bssd_fd} shows the architecture of the BSSD-FD network. The left branch performs beamforming and dereverberation, and the right branch outputs an embedding vector per utterance.

\subsection{Speaker Separation}

The STFT layer transforms the multi-channel input mixture to the frequency domain using Eq. (\ref{eq:STFT}). The STFT produces $L$ time frames and $K$ frequency bins per frame. Source separation is based on the DOA index $\hat{d}$ from Algorithm \ref{alg:localization}, which is a scalar. The \emph{Adaption} layer uses this input to modify the IPDs of the multi-channel input mixture in frequency domain, i.e.

\begin{equation}
\vm{\tilde{Z}}(l,k) = \Big( \vm{U}(k) \vm{V}(\hat{d},k) \Big)^* \odot \Big( \vm{U}(k) \vm{Z}(l,k) \Big),
\label{eq:analytic_adaption}
\end{equation}

\noindent where $^*$ denotes complex conjugation, and $\odot$ element-wise multiplication. The adaption layer performs two tasks: (i) It whitens the input signal $\vm{Z}(l,k)$ as shown in Eq. (\ref{eq:GCC-PHAT-whitened}). (ii) It subtracts the phase of the whitened DOA vector $\vm{V}(\hat{d},k)$ from the phase of the whitened input signal. This operation will align the phase of $\vm{\tilde{Z}}(l,k)$ such that the IPDs for the direction of $\vm{V}(\hat{d})$ are zero. Consequently, signals originating from this direction (i.e. the desired signal) can be identified by a small IPD, whereas all other signal components (i.e. interfering speakers) will have large IPDs. Hence, the NN sees the desired signal always at the same spatial location, enabling it to distinguish between the desired and unwanted signal components. Consequently, the NN extracts the speaker towards the direction of $\vm{V}(\hat{d})$. We refer to Eq. (\ref{eq:analytic_adaption}) as the \emph{Analytic Adaption} (AA). Hence, this system is abbreviated as BSSD-FD-AA.

Instead of modifying the phase of the input with the DOA vector, it is also possible to modify the input directly with a set of trainable weights, i.e.

\begin{equation}
\vm{\tilde{Z}}(l,k) = \vm{A}(\hat{d},k) \vm{Z}(l,k),
\label{eq:statistic_adaption}
\end{equation}

\noindent where $\vm{A}(\hat{d},k)$ is a complex-valued matrix of shape $M \times M$. It allows to scale, shift and mix the $M$ channels of the complex-valued inputs $\vm{Z}(l,k)$ freely. Note that the DOA index $\hat{d}$ selects the location from which we want to extract the desired speech signal. Hence, during training, all possible $D$ DOA locations must be presented to the NN to train all complex-valued weights in the tensor $\vm{A}$. Eq. (\ref{eq:statistic_adaption}) can be implemented as a complex-valued linear layer \cite{Pfei:may2019}. We refer to Eq. (\ref{eq:statistic_adaption}) as \emph{Statistic Adaption} (SA). Hence, this system is abbreviated as BSSD-FD-SA.

\subsection{Beamforming and Dereverberation}

The structure of the left branch of the NN in Figure \ref{fig:bssd_fd} resembles a traditional \emph{filter-and-sum} beamformer, which can be written as:

\begin{equation}
Y(l,k) = \vm{W}^T(l,k) \vm{\tilde{Z}}(l,k),
\label{eq:filter_and_sum_bf}
\end{equation}

\noindent where $Y(l,k)$ denotes the beamformed output in frequency domain, and $\vm{W}(l,k)$ are the beamforming filters. The inner vector product of Eq. (\ref{eq:filter_and_sum_bf}) is computed before the inverse STFT layer in Figure \ref{fig:bssd_fd}. The NN estimates the weights $\vm{W}(l,k)$ solely from the spatial information in $\vm{\tilde{Z}}(l,k)$, which is obtained by the \emph{Norm} layer. In particular, this layer normalizes the magnitude of the $M$ dimensional input vector $\vm{\tilde{Z}}(l,k)$ to 1, and aligns its phase to the first microphone, i.e.

\begin{equation}
\vm{v}_{\tilde{Z}}(l,k) = \frac{ \vm{\tilde{Z}}(l,k) \cdot \tilde{Z}^*(l,k,m=1) }{ |\vm{\tilde{Z}}(l,k) \cdot \tilde{Z}^*(l,k,m=1)| }.
\label{eq:norm_layer}
\end{equation}

\noindent Then, a bidirectional LSTM layer creates a latent space of $H$ neurons, followed by a dense layer with a complex-valued tanh activation function \cite{Pfei:may2019}. A linear layer outputs a set of unconstrained filter weights $\vm{W}(l,k) \in \mathbb{C}$ to calculate the enhanced output $Y(l,k)$ as shown in Eq. (\ref{eq:filter_and_sum_bf}). After the inverse STFT layer, we obtain the enhanced time-domain signal $y(t)$.

By using a neural beamformer, the design goal is not limited to MVDR constraints or similar concepts \cite{Pfei:may2019}. In fact, we can also include a dereverberation objective by using an appropriate loss function for the NN. In particular, we use the negative SI-SDR \cite{Roux:2019} between the output $y(t)$, and a clean anechoic reference utterance $r(t)$, i.e.

\begin{equation}
\mathcal{L}_{\text{SI-SDR}} = - 10log_{10} \Big( \frac{|\alpha r(t)|_2^2}{|\alpha r(t) - y(t)|_2^2} \Big),
\label{eq:si-sdr}
\end{equation}

\noindent where $\alpha = \frac{y(t)^T r(t)}{r(t)^T r(t)}$. We use $r(t) = s_c(t)$ from Eq. (\ref{eq_soundsource}) as anechoic reference signal.

\subsection{Speaker Identification}

The right branch of the NN in Figure \ref{fig:bssd_fd} extracts an embedding vector $\vm{e}$ to identify the speaker in the enhanced output signal $Y(l,k)$. The embedding vector maps the utterance into a feature space where distances correspond to speaker similarity \cite{Snyder:2016}. Therefore, the NN must be agnostic to the spoken text, and only rely on the speaker characteristics embedded in the waveform. We use the log-power spectral density $\log{|Y(l,k)|^2}$ as input features. Then, a series of 6 convolutional layers with a filter length of 10, and increasing dilation factors of (1,2,4,8,16,32) frames create a latent space of $L \times E$ dimensional embeddings. We use a softplus\footnote{The softplus activation function is defined as $f(z) = \log{1+\exp{z}}$.} activation function and layer normalization after each convolutional layer. A skip connection is added between every two convolutional layers. The $L$ time frames are averaged to obtain a single, $E$ dimensional embedding for the whole utterance (AvgPool layer). The linear layer at the end of the stack outputs the unconstrained embedding vector $\vm{e}$.

We want to identify an open set of speakers, i.e. we need to be able to compare two utterances and determine whether they belong to the same speaker. Therefore we employ the triplet loss \cite{Schroff:2015}, which has been successfully used for speaker identification and diarization tasks \cite{Li:2017, Wang:sep2017, Song:2018, Zhang:2018}. The goal of the triplet loss is to ensure that two utterances from the same speaker have their embeddings close together in the embedding space, and two examples from different speakers have their embeddings farther away by some margin $\beta$. In other words, we want the embeddings of the same speaker to form clusters, and these clusters must be separated by the margin, i.e.

\begin{equation}
\mathcal{L}_{\text{TL}} = \sum_{B^3} \Big[ |\vm{e}_a-\vm{e}_p|_2 - |\vm{e}_a-\vm{e}_n|_2 + \beta \Big]_+,
\label{eq:triplet_loss}
\end{equation}

\noindent where the embedding $\vm{e}_a$ denotes an \emph{anchor}, $\vm{e}_p$ is an embedding from the same speaker as the anchor (positive example), and $\vm{e}_n$ is an embedding from a different speaker (negative example). In a batch of $B$ utterances, there can be as many as $B^3$ triplets. It is therefore crucial to only select a subset of valid triplets, where the positive example is from the same speaker as the anchor, and the negative example belongs to a different speaker. Further, we only need to consider triplets where the loss $\mathcal{L}_{\text{TL}}$ is actually greater than zero. To select relevant triplets, we utilize \emph{Hard Triplet Mining} \cite{Hermans:2017}, where we select the hardest positive and negative example per anchor. In particular, we randomly select $P$ utterances from $B$ speakers, where we determine the largest distance $|\vm{e}_a-\vm{e}_p|_2$ between an anchor and a positive example within the $P$ utterances per speaker, and the smallest distance $|\vm{e}_a-\vm{e}_n|_2$ between an anchor and a negative example from the $P(B-1)$ remaining utterances. More formally, this procedure can be written as:

\begin{equation}
\begin{aligned}
\mathcal{L}_{\text{TL-HTM}} = \frac{1}{B \cdot P} \sum_{i=1}^B \sum_{a=1}^P \Big[ \beta &+ \underset{p=1 \ldots P}{\text{max}} \Big( |\vm{e}_a^i-\vm{e}_p^i|_2 \Big) \\ &- \underset{\substack{j=1 \ldots B \\ n= 1 \ldots P \\ i \neq j}}{\text{min}} \Big( |\vm{e}_a^i-\vm{e}_n^j|_2 \Big) \Big]_+.
\end{aligned}
\label{eq:hard_triplet_mining}
\end{equation}

\noindent When the batch size $P \cdot B$ is small, the embeddings may collapse into a single point during training \cite{Zhang:aug2017}. To avoid this, we propose to minimize the cross-entropy between embeddings of different speakers as follows:

\begin{equation}
\mathcal{L}_{\text{TL-CE}} = \frac{-1}{(B^2-B) P^2} \sum_{a=1}^B \sum_{\substack{n=1 \\ n \neq a}}^B \sum_{i=1}^P \sum_{j=1}^P \log{ |( \tilde{\vm{e}}_a^i)^T \tilde{\vm{e}}_n^j|^2 }, 
\label{eq:ce_regularization}
\end{equation}

\noindent where $\tilde{\vm{e}} = \frac{\vm{e}}{|\vm{e}|_2}$ is the magnitude-normalized embedding vector $\vm{e}$. This regularization ensures that the embeddings $\vm{e}_a$ and $\vm{e}_n$ will be different. The overall cost function for the entire BSSD-FD architecture is then defined as:

\begin{equation}
\mathcal{L}_{\text{BSSD-FD}} = \mathcal{L}_{\text{SI-SDR}} + \lambda_1 \mathcal{L}_{\text{TL-HTM}} + \lambda_2 \mathcal{L}_{\text{TL-CE}}, 
\label{eq:bssd_fd_loss}
\end{equation}

\noindent where $\lambda_1$ and $\lambda_2$ are weights for the individual terms.

\subsection{Distance Measure}

In order to determine whether two embeddings $\vm{e}_1$ and $\vm{e}_2$ belong to the same speaker, we use the euclidian distance $|\vm{e}_1 - \vm{e}_2|_2$. If the distance falls below a certain threshold $\delta$, we consider the two embeddings to belong to the same speaker. If it exceeds the threshold, the speakers are considered different. Hence, two types of errors exist: (i) A false positive is triggered when two embeddings from two different speakers are incorrectly classified as belonging to the same speaker, which we measure using the False Acceptance Rate (FAR), i.e.

\begin{equation}
\text{FAR}(\delta) = \frac{1}{(B^2-B) P^2} \sum_{a=1}^B \sum_{\substack{n=1 \\ n \neq a}}^B \sum_{i=1}^P \sum_{j=1}^P \mathds{1} \Big( |\vm{e}_a^i-\vm{e}_n^j|_2 < \delta \Big),
\label{eq:FAR}
\end{equation}

\noindent where $\mathds{1}(x)$ denotes an indicator function, i.e.

\begin{equation}
\mathds{1}(x) = \begin{cases}
    1, & \text{if condition } $x$ \text{ is true}.\\
    0, & \text{otherwise}.
  \end{cases}  
\end{equation}

\noindent (ii) A false negative is triggered when two embeddings from the same speaker are classified as belonging to different speakers, which we measure using the False Rejection Rate (FRR), i.e.

\begin{equation}
\text{FRR}(\delta) = \frac{1}{B(P^2-P)} \sum_{\substack{a=1 \\ p=a}}^B \sum_{i=1}^P \sum_{\substack{j=1 \\ j \neq i}}^P \mathds{1} \Big( |\vm{e}_a^i-\vm{e}_p^j|_2 > \delta \Big).
\label{eq:FRR}
\end{equation}

\noindent The FAR is positively correlated to the decision threshold $\delta$, and the FRR is correlated negatively. The value at which the FAR and FRR are equal, is known as the Equal Error Rate (EER). It is determined by:

\begin{equation}
\begin{aligned}
\hat{\delta} &= \underset{\delta}{\text{argmin}} |\text{FAR}(\delta)-\text{FRR}(\delta)| \\
\text{EER} &= \text{FAR}(\hat{\delta}) = \text{FRR}(\hat{\delta}),
\end{aligned}
\label{eq:EER}
\end{equation}

\noindent where $\hat{\delta}$ is considered the optimal threshold belonging to the EER.

\section{BSSD Network - Time Domain}
\label{sec:bssd_td}

\begin{figure}[!ht]
\centering
\includegraphics[width=0.9\linewidth]{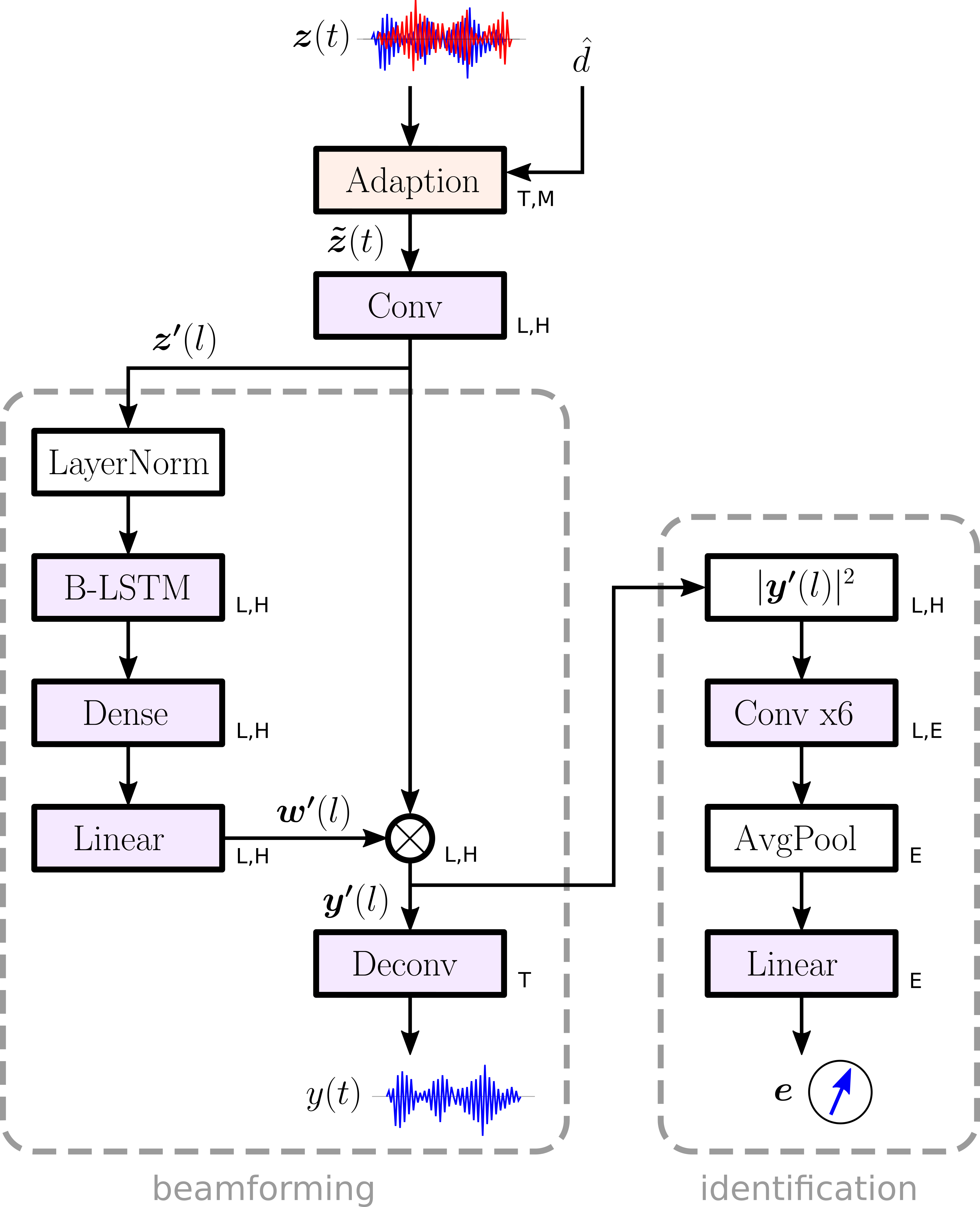}
\caption{Layers of the time-domain BSSD-TD network. The left branch performs beamforming and dereverberation, the right branch assigns an embedding vector to the enhanced output signal $y(t)$.}
\label{fig:bssd_td}
\end{figure}

With the recent success of time-domain speech separation algorithms \cite{Stoller:2018, Luo:2018, Luo:2019, Chang:2019}, we also formulate a time-domain variant of our BSSD network. The beamforming and dereverberation operations can be formulated analogously to the frequency domain, but the NN will train faster when using real-valued data. Figure \ref{fig:bssd_td} shows the architecture of the BSSD-TD network. Similar to the frequency-domain variant, the left branch performs beamforming and dereverberation, and the right branch outputs an embedding vector per utterance.

\subsection{Speaker Separation}

Analogous to the frequency-domain network, source separation is based on the \emph{Adaption} layer, which uses the DOA index $\hat{d}$ from Algorithm \ref{alg:localization} to modify the ITD of the input signal $\vm{z}(t)$. By rearranging Eq. (\ref{eq:analytic_adaption}), we can formulate an identical operation in time-domain, i.e.

\begin{equation}
\begin{aligned}
\tilde{Z}(l,k,m) &= \big( \vm{U}^T(k,m) \vm{V}(\hat{d},k) \big)^* \cdot \big( \vm{U}^T(k,m) \vm{Z}(l,k) \big), \\
 &= \sum_{i=1}^M \vm{U}^H(k,m) \vm{V}^*(\hat{d},k) U(k,m,i) \cdot Z(l,k,i), \\
 &= \sum_{i=1}^M V'(\hat{d},k,m,i) \cdot Z(l,k,i),
\end{aligned}
\end{equation}

\noindent where we can identify the convolutional kernel $V'(\hat{d},k,m,i)$ in frequency domain. We can see from Eq. (\ref{eq:doa_vector}), that the DOA $\vm{V}(\hat{d},k)$ resembles $M$ sinc pulses with a positive time-delay $\tau_{d,m}$, and Eq. (\ref{eq_whitening}) shows that the elements of the whitening matrix $U(k,m,i)$ are real-valued. Therefore, $V'(\hat{d},k,m,i)$ will be a causal IIR filter in time domain \cite{bible1}, which we truncate to $T_A$ samples to obtain the FIR filter $v'(\hat{d},t_A,m,i)$ by using the inverse FFT. This allows to formulate the time-domain adaption layer as:

\begin{equation}
\tilde{z}(t,m) = \sum_{i=1}^M z(t,i) \circledast v'(\hat{d},t_A,m,i),
\label{eq:analytic_adaption_td}
\end{equation}

\noindent which can be implemented using a single convolution layer. Similar to the frequency-domain adaption layer, Eq. (\ref{eq:analytic_adaption_td}) synchronizes the ITD to be zero for signals originating from the direction of $\vm{V}(\hat{d})$, i.e. the desired signal. The subsequent NN sees the desired signal always at the same spatial location, which makes it easier to distinguish between the desired and unwanted signal components. Consequently, the NN extracts the speaker towards the direction of $\vm{V}(\hat{d})$. We refer to Eq. (\ref{eq:analytic_adaption_td}) as \emph{Analytic Adaption} (AA). Hence, this system is abbreviated as BSSD-TD-AA.

Instead of modifying the ITDs of the input signal with the DOA vector, it is also possible to replace the fixed convolutional kernels $\tilde{v}(\hat{d},t_A,m,i)$ with a set of trainable weights, i.e.

\begin{equation}
\tilde{z}(t,m) = \sum_{i=1}^M z(t,i) \circledast a(\hat{d},t_A,m,i),
\label{eq:statistic_adaption_td}
\end{equation}

\noindent where $\vm{a}$ is a tensor of shape $(D,T_A,M,M)$, and $T_A$ is the filter length of the learnable convolution kernels. This allows to scale, shift and mix the $M$ channels of the input signal $\vm{z}(t)$ freely. Note that the DOA index $\hat{d}$ provides the location from which we want to extract the desired speech signal. Hence, during training, all $D$ possible DOA locations must be presented to the NN to train the weights $\vm{a}$. Note that Eq. (\ref{eq:statistic_adaption_td}) can be implemented as a convolution layer. We refer to Eq. (\ref{eq:statistic_adaption_td}) as \emph{Statistic Adaption} (SA). Hence, this system is abbreviated as BSSD-TD-SA.

\subsection{Beamforming and Dereverberation}

The structure of the left branch of the NN in Figure \ref{fig:bssd_td} resembles a time-domain  beamformer \cite{micarray1}, where the first convolution layer right after the adaption layer transforms the time-domain input $\vm{\tilde{z}(t)}$ into a latent space $z'(l,h)$ with $L$ frames and $H$ filters. The stride of this convolution layer is set to $\frac{H}{4}$, and the activation function is linear. 

Similar to the frequency-domain beamformer, filtering is performed in latent space. The beamforming weights $w'(l,h)$ are estimated from the spatial information embedded in $z'(l,h)$, using layer normalization, followed by a bidirectional LSTM layer, a dense layer with tanh activation, and a linear layer. The linear layer allows the NN to freely choose the amplitude and phase of the beamforming weights. The enhanced output $y'(l,h)$ is obtained by

\begin{equation}
  y'(l,h) = w'(l,h) \odot z'(l,h),
\label{eq:filter_and_sum_bf_td}
\end{equation}

\noindent where all variables are of shape $L \times H$. Finally, a deconvolution layer with a linear activation function produces the enhanced time-domain signal $y(t)$. Analogous to the BSSD-FD architecture, we use the negative SI-SDR from Eq. (\ref{eq:si-sdr}) between the output $y(t)$, and a clean anechoic reference utterance $r(t)$.

\subsection{Speaker Identification}

The right branch of the NN in Figure \ref{fig:bssd_td} extracts an embedding vector $\vm{e}$ to identify the speaker in the enhanced output signal $y(t)$. The NN is identical to the BSSD-FD architecture, except for the input layer which uses the enhanced signal $y'(l,h)$ as input features. Identically to Eq. (\ref{eq:bssd_fd_loss}), the overall cost function for the entire BSSD-TD architecture is defined analogously to the BSSD-FD architecture given in Eq. \ref{eq:bssd_fd_loss}.

\section{Block Online Processing}
\label{sec:block_online_processing}

\begin{figure}[!ht]
\centering
\includegraphics[width=0.8\linewidth]{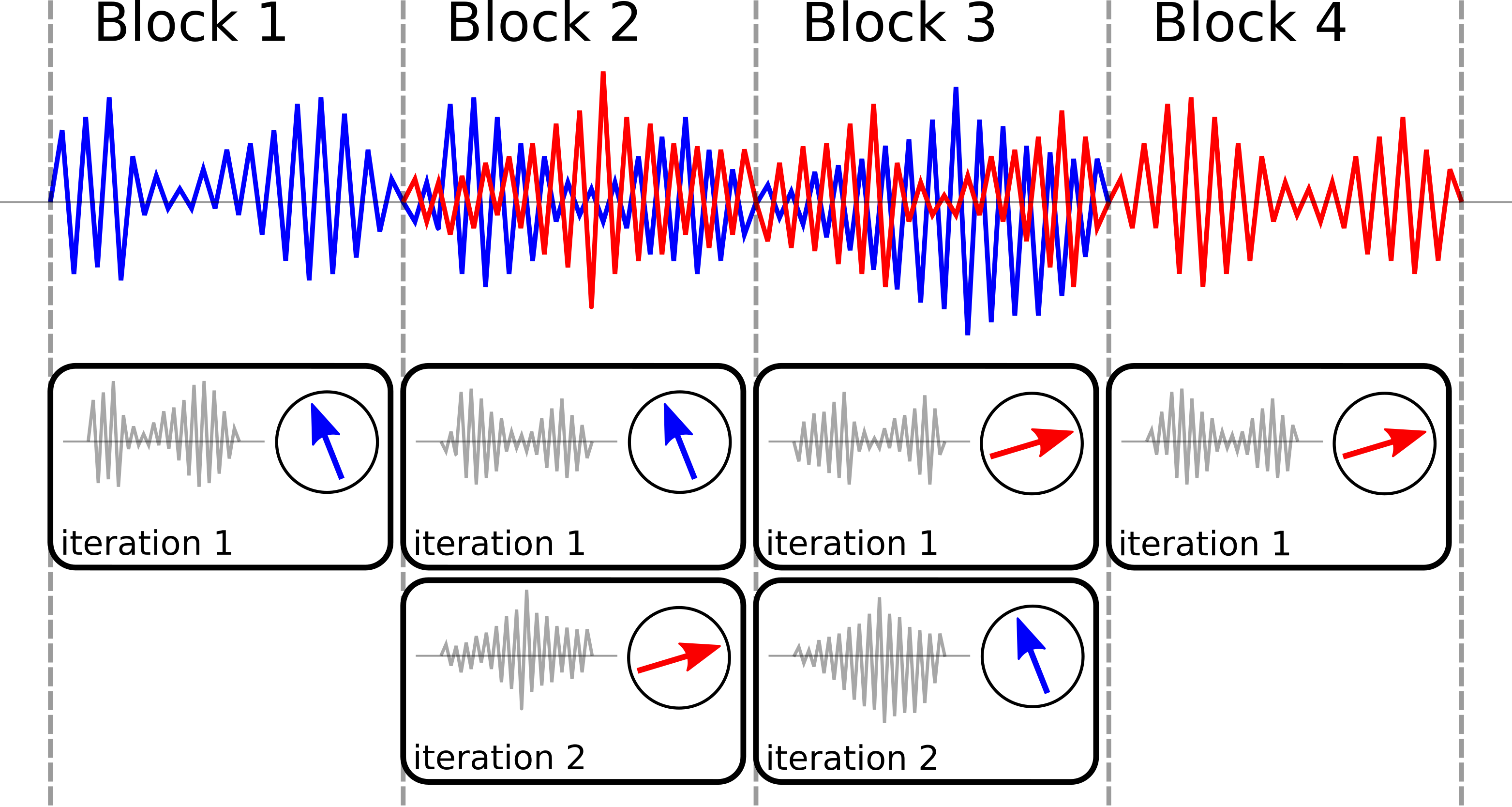}
\caption{Block-online processing mode of the BSSD system, showing a mixture of $C=2$ speakers being split into 4 blocks. Each block is processed separately using Algorithm \ref{alg:localization}.}
\label{fig:block_online_processing}
\end{figure}

For realtime applications, it is possible to use the BSSD system in block-online mode. We split the input mixture $\vm{z}(t)$ into blocks of equal length. Each block $b$ is iteratively processed using Algorithm \ref{alg:localization}. It returns the DOA index $\hat{d}$, a list $\mathcal{Y}_b$ of extracted speakers $y(t)$, and a list $\mathcal{E}_b$ of speaker embeddings $\vm{e}$. Figure \ref{fig:block_online_processing} illustrates the block-online processing scheme of the BSSD system for $C=2$ speakers and 4 blocks. Note that the speakers may change their position from block to block, i.e. they may walk around, switch places, etc. Therefore, both the DOA and speaker embedding are extracted for each source in each block. The process of assigning each extracted source block to the same speaker identity is referred to as \emph{diarization} \cite{Sell:2018, Yu:jul2017}, by using Algorithm \ref{alg:block_online}.

\begin{algorithm}[h]
\begin{algorithmic}[1]
  \State $\mathcal{Y} \gets []$
  \State $\mathcal{E} \gets []$

  \For{$\textbf{all blocks } b$}
    \For{$c = 1:\textbf{length}(\mathcal{E}_b)$}
	
	  \If{$\texttt{min}(|\mathcal{E}-\mathcal{E}_b(c)|_2) > \hat{\delta}$}
        \State $\mathcal{E} \texttt{.append}(\mathcal{E}_b(c))$
      \Else
        \State $i \gets \texttt{argmin}(|\mathcal{E}-\mathcal{E}_b(c)|_2))$
        \State $\mathcal{Y}(i) \texttt{.append}(\mathcal{Y}_b(c))$
      \EndIf
      
    \EndFor
  \EndFor
  
\end{algorithmic}
\caption{Diarization in block-online mode.}
\label{alg:block_online}
\end{algorithm}

First, we initialize empty lists for all speakers $\mathcal{Y}$ and all embeddings $\mathcal{E}$. Then, we iterate over all blocks $b$, where Algorithm \ref{alg:localization} is executed for each block. It returns a list $\mathcal{Y}_b$ of extracted speakers and a list $\mathcal{E}_b$ of speaker embeddings for that block. Next, we iterate over each extracted source $c$ within that block, and we compare the distance of the embedding $\mathcal{E}_b(c)$ against all embeddings $\mathcal{E}$. If the threshold $\hat{\delta}$ (see Eq. \ref{eq:EER}) is exceeded, we have found a new speaker. In that case, this speaker is added to the list of known embeddings $\mathcal{E}$. Otherwise, we have found an utterance belonging to a known embedding. In that case,  we determine the index $i$ of that embedding, and append the source $\mathcal{Y}_b(c)$ to the speaker at position $\mathcal{Y}(i)$. To preserve the time alignment of each extracted speaker, we append a block of silence to each source in $\mathcal{Y}$ that did not receive an update. This may happen if a speaker is silent during block $b$.

A trade-off has to be made when choosing the block length $T_B$. If it is too large, short utterances followed by periods of silence might not get detected. If it is too small, the predicted embeddings may be inaccurate, causing Algorithm \ref{alg:block_online} to assign the sources $\mathcal{Y}_b(c)$ to the wrong speaker. We examine this behavior in our experiments.

\section{RIR Recordings}
\label{sec:rir_recordings}

We use both recorded and simulated RIRs to generate spatialized recordings with Eq. (\ref{eq_soundsource}). Real RIRs are obtained through multi-channel room impulse response measurements, and simulated RIRs are obtained by the Image Source Method (ISM) \cite{Wang:2018, Habets2007}.

\subsection{Real RIRs}

\begin{figure}[!ht]
\centering
\includegraphics[width=0.5\linewidth]{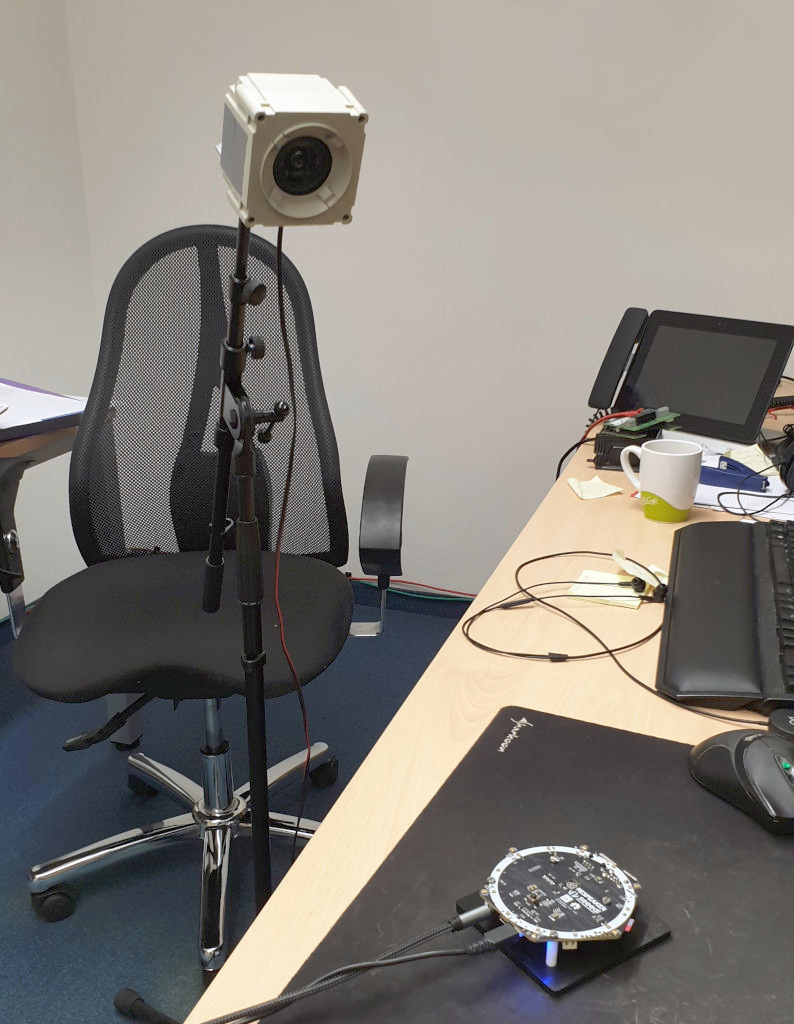}
\caption{RIR recording setup using a 5W measurement loudspeaker and a 6-channel microphone array \cite{respeaker}.}
\label{fig:rir_recording}
\end{figure}

To obtain realistic RIRs, we use a circular microphone array with $M=6$ channels and a diameter of $92.6mm$ \cite{respeaker}, and a 5W measurement loudspeaker. To drive the loudspeaker from a Linux-based PC with ALSA \cite{ALSA}, we use the PlayRec Python module \cite{sounddevice}, which simultaneously plays and records audio from a sound card. We use an exponential chirp with a duration of 5s sweeping from $24kHz$ down to $20Hz$ as excitation signal \cite{roomacoustics1}. However, we only use a bandwidth of $8kHz$ for our experiments. We recorded 120 6-channel RIRs in 24 different, fully furnished office rooms with a reverberation time $RT_{60} \in [200 \ldots 900]ms$. The distance from the loudspeaker to the microphone array was varied from $1m \ldots 3m$, and the direction was chosen randomly. Figure \ref{fig:rir_recording} shows the recording setup. We augmented the number of RIR recordings to 720 by virtually rotating the array by $6 \times 60^{\circ}$, i.e. shifting the microphone channels. 

\subsection{Simulated RIRs}

To obtain simulated RIRs, we further generated 720 artificial RIRs for the same array geometry with 6 channels, but with a shorter  reverberation time $RT_{60} \in [200 \ldots 400]ms$, which is randomly chosen. The room is modeled as a simple rectangular shoebox with random dimensions ranging from $3m \ldots 6m$, where the microphone array and the sound sources are placed randomly. RIR generation was done using the Image Source Method \cite{Wang:2018, Habets2007} using \emph{Pyroomacoustics} \cite{pyroomacoustics}.

\section{Experiments}

\subsection{Experimental Setup}

\subsubsection{Speech mixtures}

We use the WSJ0 speech database which contains 12776 utterances from 101 different speakers for training, and 5895 utterances from 18 different speakers for testing. To generate mixtures, we use the wsj0-2mix from \cite{Hershey:2016}, which we extended to 3 and 4 speakers. To generate reverberated, multi-channel mixtures from Eq. (\ref{eq_soundsource}), we convolve the monaural signals with both the \emph{real} and \emph{simulated} RIRs, as described in Section \ref{sec:rir_recordings}. 
All recordings use a sample rate of $f_s=16kHz$.

\subsubsection{DOA bases}

We use $D = 100$ DOA bases, which are equally distributed on a sphere, as shown in Figure \ref{fig:doa_sphere}. This provides an average spatial resolution of about $13.82^{\circ}$, which equates to two persons standing right next to each other at a distance of 1m. Note that the number of DOA vectors $D$ can be changed without re-training the NN. Clearly, we want to use a different DOA index $d_c \in [1 \ldots D]$ for each source $s_c$ in the input mixture $\vm{z}(t)$. To achieve this, we randomly select a RIR $\vm{h}_c(t)$ belonging to a the DOA index $d_c$ using Eq. (\ref{eq:GCC-PHAT_doa}). From the 720 \emph{real} and \emph{simulated} RIRs available, 640 are used for training, and 80 for testing.

\subsubsection{BSSD-FD system}

For the BSSD-FD network in Figure \ref{fig:bssd_fd}, we use a FFT length of 1024 samples, and an overlap of 75\%. This results in $K=513$ frequency bins. Further, we have $M=6$ microphones as determined by the RIR recordings. The \emph{beamforming} branch uses $H=500$ neurons to create the beamforming weights $\vm{W}(l,k)$, and to predict the enhanced signal $Y(l,k)$. The \emph{identification} branch uses an embedding dimension of $E=100$ to predict the speaker embeddings $\vm{e}$.

\subsubsection{BSSD-TD system}

For the BSSD-TD network in Figure \ref{fig:bssd_td}, we use a filter length of $T_A=100$ samples for the filter kernels in the adaption layer in Eq. (\ref{eq:analytic_adaption_td}) and (\ref{eq:statistic_adaption_td}). The first convolutional layer uses a filter length of 200 samples and a stride of 50 samples to create a latent space of $H=500$ neurons. The \emph{beamforming} branch predicts the beamforming weights $\vm{w'}(l)$ and the enhanced signal $\vm{y'}(l)$ in latent space. This signal is transformed back to time domain using the deconvolution layer, which uses a filter length of 200 samples, a stride of 50 samples, and overlap-add to produce the enhanced signal $y(t)$. The \emph{identification} branch uses an embedding dimension of $E=100$ to predict the speaker embeddings $\vm{e}$.

\subsection{Related Systems}

To compare our BSSD system against other state-of-the art speech separation algorithms, we evaluate Conv-TasNet \cite{Luo:2019} and PIT with spatial features \cite{Wang:2018}. To account for dereverberation, we apply the WPE algorithm from  \cite{Yoshioka:2009} prior to Conv-TasNet.

\subsubsection{Conv-TasNet}

Conv-TasNet separates 2 speakers in time domain. It operates on chunks of 4s of audio, where it separates the two speakers in a latent space by using a speech mask $\in [0 \ldots 1]$. The mask is obtained from a series of convolutions. The system operates on single-channel inputs, therefore we only use one output of the WPE algorithm. We use the implementation of \cite{Luo:2019}.

\subsubsection{Spatial PIT}

Spatial PIT separates 2 speakers in frequency domain. It uses log-spectrograms and the sine and cosine of the IPDs of frequency-domain, multi-channel mixtures to predict a speech mask for each speaker \cite{Yu:2017, Wang:2018}. This mask is used to construct a frequency-domain beamformer \cite{Pfei:aug2019}. Note that there is no explicit dereverberation constraint, but the target speech mask is obtained from the anechoic reference signal $r(t)$. Hence, the beamformer will remove late echoes.

\subsection{Training}

All four variants of the BSSD network are trained on mixtures of $C=2$ sources, where each mixture $\vm{z}(t)$ is truncated to 5s length. The location (i.e. the DOA index $d$) of each source is chosen randomly for each example. We use a batch size of 60 mixtures from the 101 speakers of the WSJ0 training set. To enable efficient triplet mining with Eq. (\ref{eq:hard_triplet_mining}), we use $P=3$ different utterances from $B=20$ speakers for the first source $s_{c=1}(t)$ of each mixture. The second source $s_{c=2}(t)$ is chosen randomly from the remaining 100 speakers from the WSJ0 training set. We use the clean first source as as reference utterance, i.e. $r(t) = s_{c=1}(t)$. The ground truth DOA index $\hat{d}$ is used to train the network. We use $\lambda_1 = 10^{-2}$ and $\lambda_2 = 10^{-4}$ for the cost function in Eq. (\ref{eq:bssd_fd_loss}). This ensures that the \emph{beamforming} path is trained faster than the \emph{identification} path, as the latter depends on the former. As the combination of the different RIRs and WSJ0 utterances allows for billions of combinations, we randomly create new batches for training and validation for each epoch. Adam is used as optimizer \cite{Kingma:Jul15}, with a learning rate of $10^{-3}$. A Tensorflow implementation of the BSSD network can be found at\footnote{https://github.com/rrbluke/BSSD}.

\subsection{Testing}

We compare the frequency-domain (FD) and time-domain (TD) variants of our BSSD system, as well as the analytic adaption (AA) and statistic adaption (SA) layers introduced in Section \ref{sec:bssd_fd} and \ref{sec:bssd_td}. Further, we use the Conv-TasNet and spatial PIT as baseline systems. We test the BSSD network both in \emph{offline} mode in \emph{block-online} mode. 

\subsubsection{Offline Mode}

In offline mode, we use 5s long mixtures of $C=\{1,2,3,4\}$ speakers from the test set. To be able to test the performance of the \emph{speaker separation} and \emph{speaker identification} modules separately, we use the ground truth DOA index $\hat{d}$ as input to the BSSD network. We report the separation and dereverberation performance in terms of SI-SDR using Eq. (\ref{eq:si-sdr}). Further, we report the WER using the Google Speech-to-Text API \cite{pySpeechRecognition} to perform ASR. Speaker identification performance is reported in terms of EER on the enhanced output, using Eq. (\ref{eq:EER}).

\subsubsection{Block-online Mode}

In block-online mode, we use 20s long mixtures from the test set, which we divide into $N_b$ blocks of $T_B=\{1,2.5,5\}s$ length. Each block is processed by Algorithm \ref{alg:localization}, which outputs the DOA index $\hat{d}$, a list of extracted signals $\mathcal{Y}_b$, and a list of speaker embeddings $\mathcal{E}_b$ for each block $b$. Then, Algorithm \ref{alg:block_online} is used to assign the extracted utterances of each block to the same speaker. This solves the speaker permutation problem. We report the SI-SDR, WER and the Block Error Rate (BER) for the extracted speakers. The BER indicates the percentage of falsely assigned blocks due to erroneous embeddings. It is determined by comparing the speaker embedding of the reference utterance $r_c(t)$ against the extracted chunks $y_{b,c}(t)$ for each speaker $c$, i.e.

\begin{equation}
  \text{BER} = \frac{1}{C \cdot N_b} \sum_{c=1}^C \sum_{b=1}^{N_b} \mathds{1} \Big( |\vm{e}_{r,c}-\vm{e}_{b,c}|_2 > \hat{\delta} \Big)
\label{eq:ber}
\end{equation}

\section{Results}

\subsection{Source Localization}

\begin{figure}[!ht]
\centering

\begin{minipage}{0.49\linewidth}
  \begin{center}
  (a)
  \includegraphics[width=1.05\linewidth]{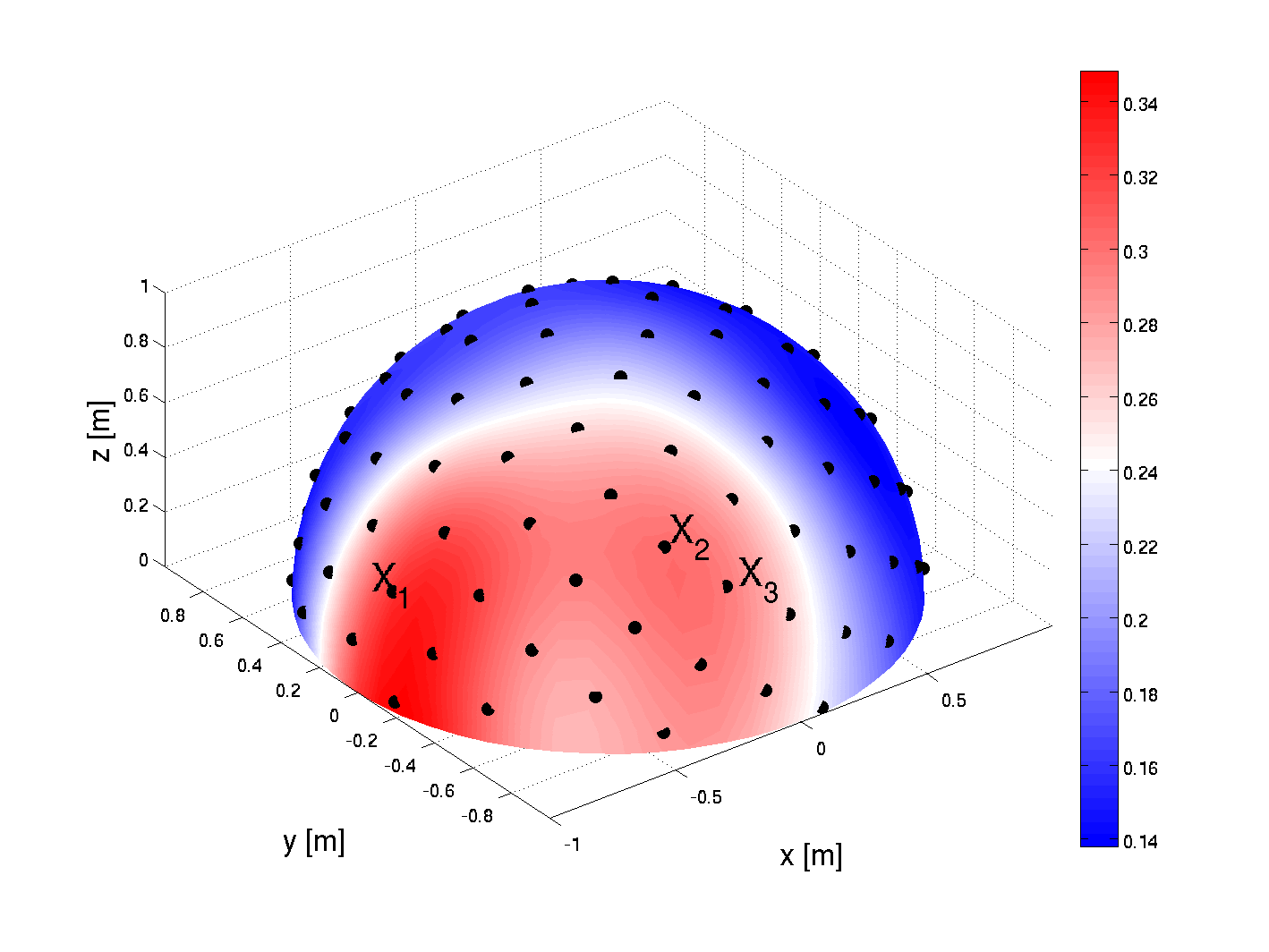}
  \end{center}
\end{minipage}
\begin{minipage}{0.49\linewidth}
  \begin{center}
  (b)
  \includegraphics[width=1.05\linewidth]{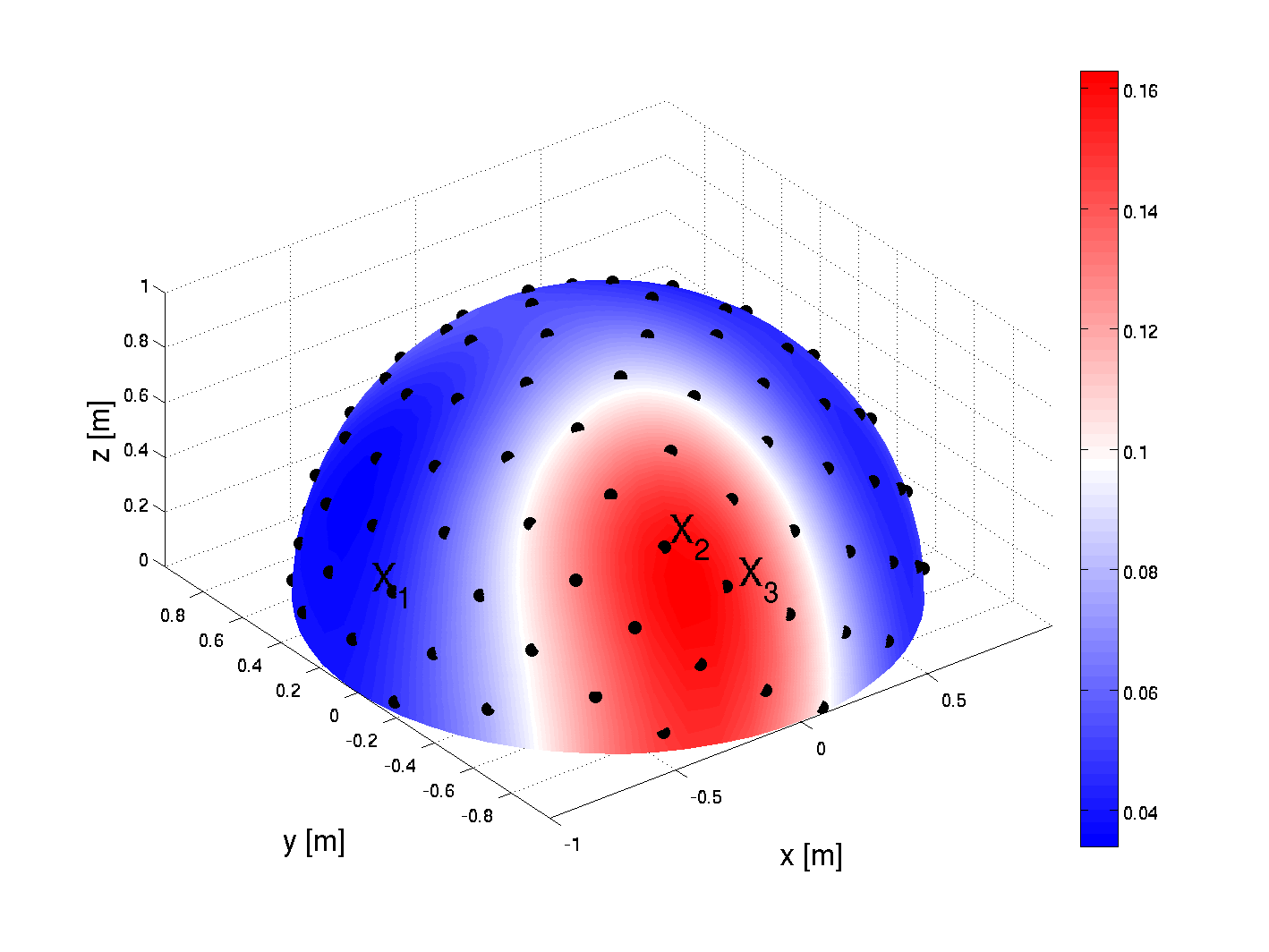}
  \end{center}
\end{minipage}
\begin{minipage}{0.49\linewidth}
  \begin{center}
  (c)
  \includegraphics[width=1.05\linewidth]{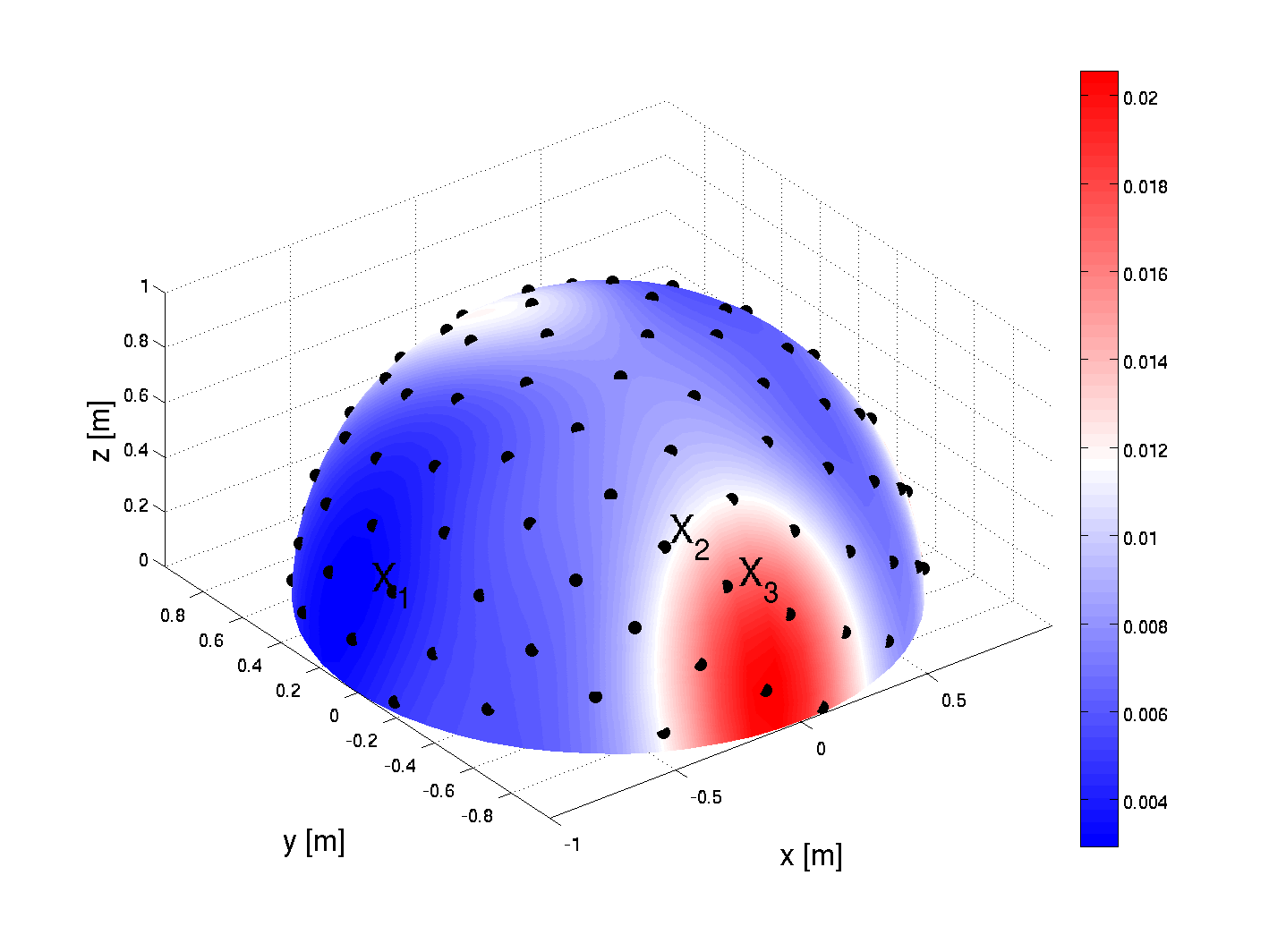}
  \end{center}
\end{minipage}
\begin{minipage}{0.49\linewidth}
  \begin{center}
  (d)
  \includegraphics[width=1.05\linewidth]{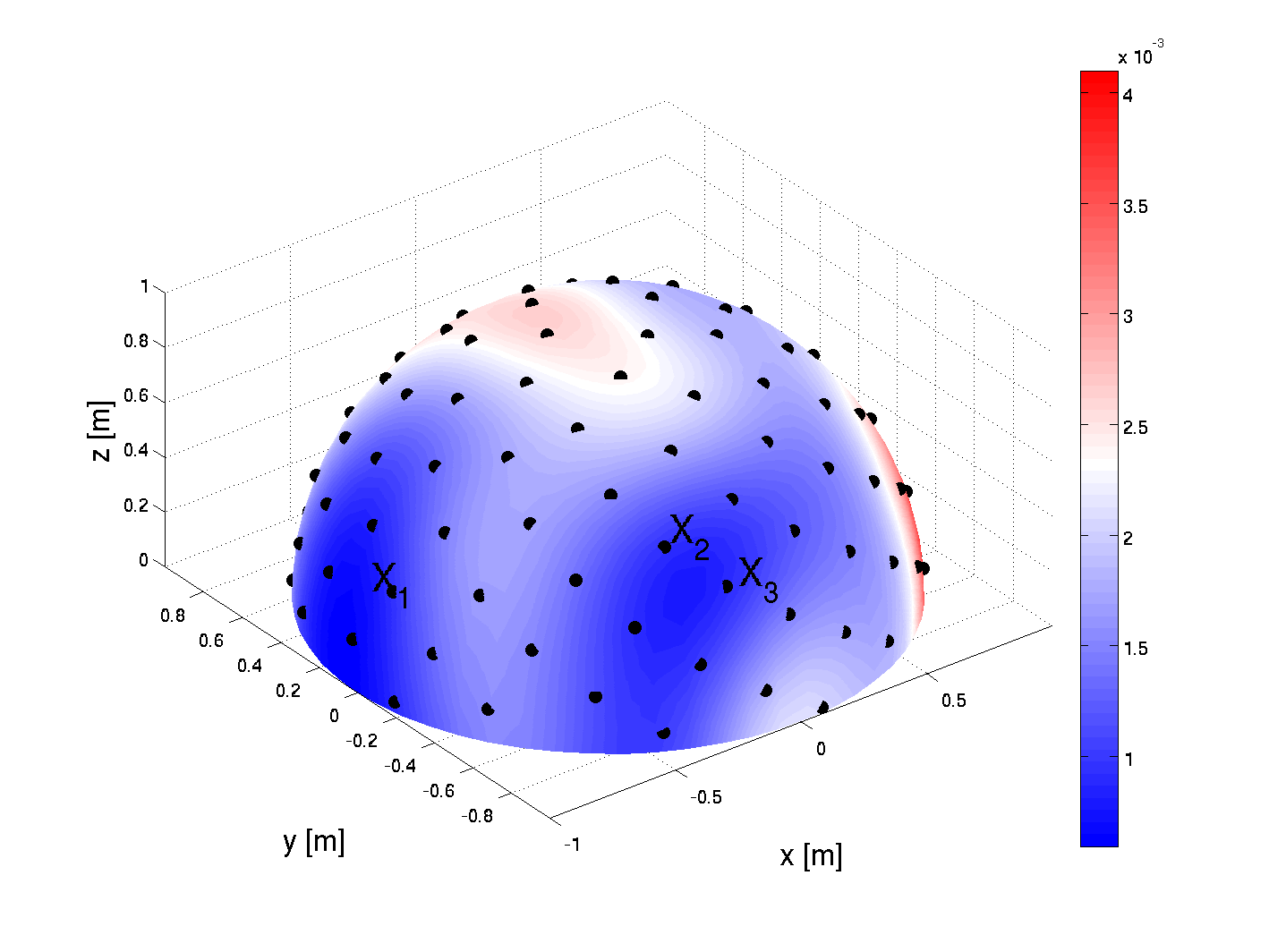}
  \end{center}
\end{minipage}

\caption{Unit sphere with $D=100$ equidistant DOA points and a circular microphone array with $M=6$ channels. Panel (a) - (d) show the spatial speech presence probability map $\gamma_W'$ during consecutive iterations of Algorithm \ref{alg:localization}.}
\label{fig:doa_sphere}
\end{figure}

First, we report the performance of the localization stage, i.e. Algorithm \ref{alg:localization}, using a mixture with $C=3$ speakers and \emph{real} RIRs. Figure \ref{fig:doa_sphere}  illustrates the $D=100$ DOA positions, arranged on a unit sphere according to Eq. (\ref{eq:doa_sphere}). The color gradient indicates the spatial speech presence probability map $\gamma_W'(d)$, which is obtained by summing Eq. (\ref{eq:GCC-PHAT-whitened}) over all frequencies $K$ and time frames $L$. The markers $X_1$, $X_2$ and $X_3$ indicate the positions of the three speakers. Panel (a) illustrates $\gamma_U(d)$ for the first iteration of Algorithm \ref{alg:localization}, panel (b) for the second iteration, panel (c) for the third iteration, and panel (d) for the fourth iteration. All three sources are localized during the first three iterations, while the fourth iteration only sees a faint reflection of one of the three sources. Therefore, the speaker embedding for this reflection will already be known, which serves as the stopping criterion of Algorithm \ref{alg:block_online}.

\subsection{Separation Performance}

\begin{figure}[!ht]
\centering

\begin{minipage}{0.95\linewidth}
  \begin{center}
  (a)
  \includegraphics[width=1.05\linewidth]{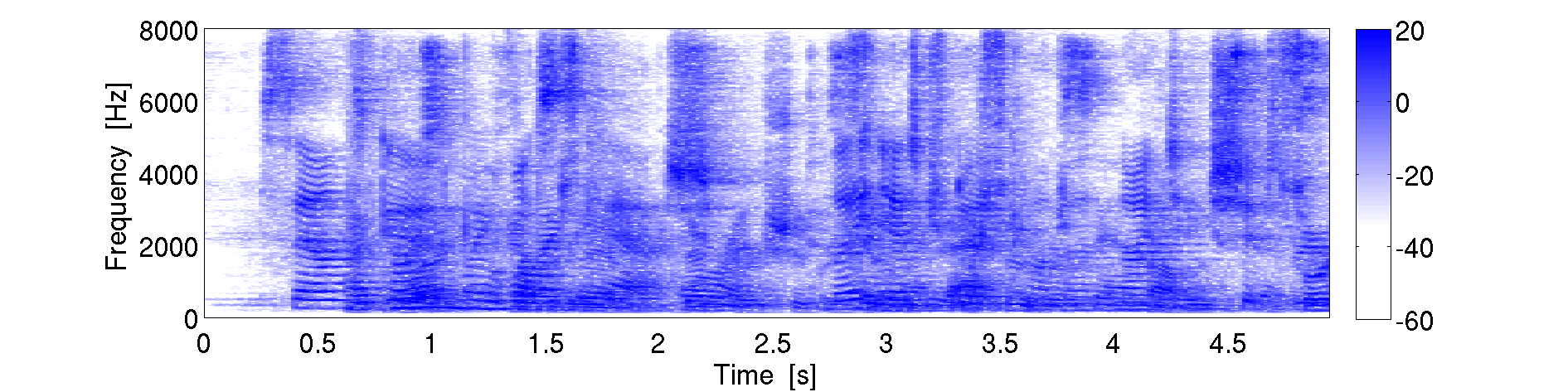}
  \end{center}
\end{minipage}
\begin{minipage}{0.95\linewidth}
  \begin{center}
  (b)
  \includegraphics[width=1.05\linewidth]{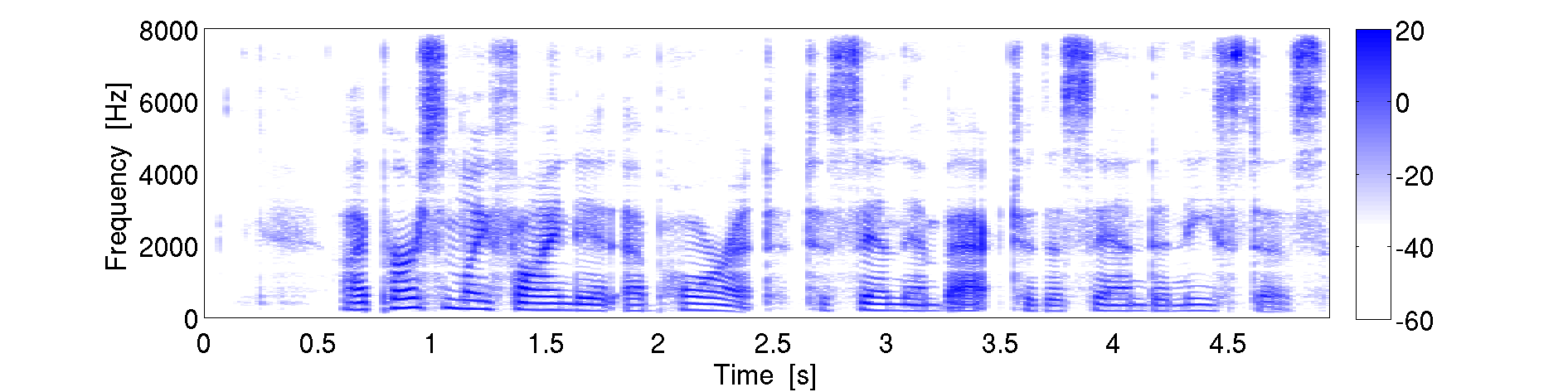}
  \end{center}
\end{minipage}
\begin{minipage}{0.95\linewidth}
  \begin{center}
  (c)
  \includegraphics[width=1.05\linewidth]{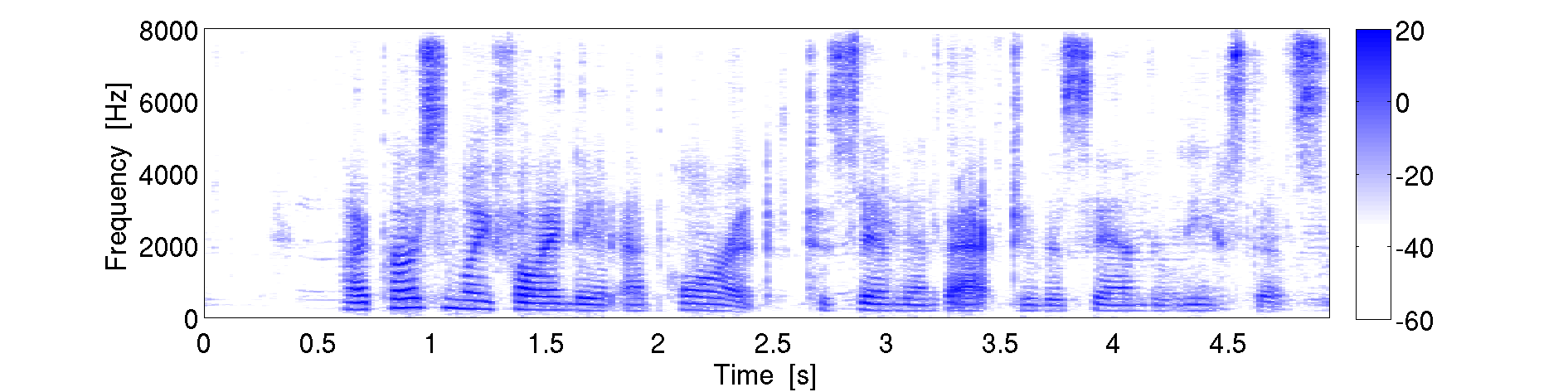}
  \end{center}
\end{minipage}
\begin{minipage}{0.95\linewidth}
  \begin{center}
  (d)
  \includegraphics[width=1.05\linewidth]{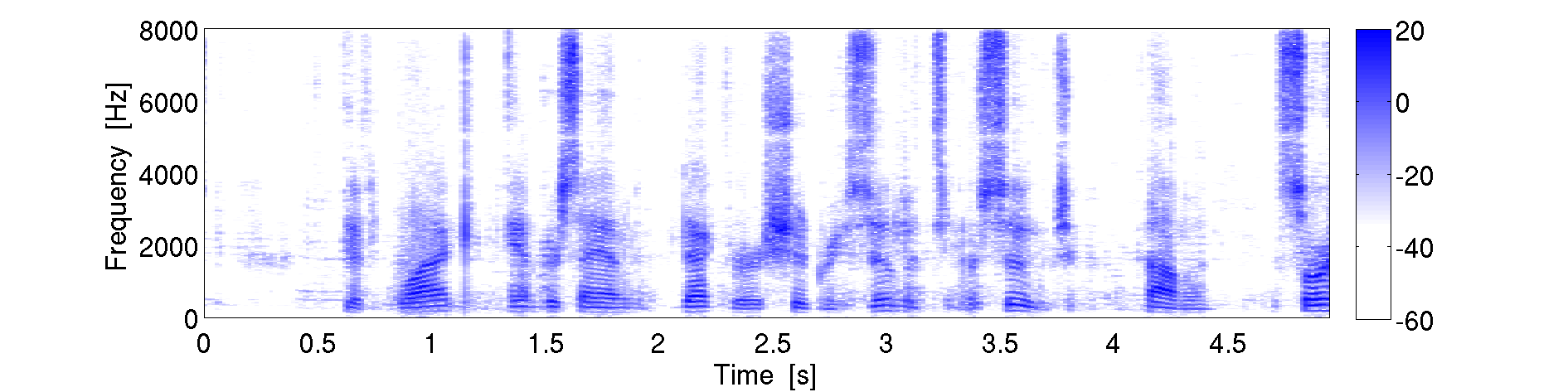}
  \end{center}
\end{minipage}
\begin{minipage}{0.95\linewidth}
  \begin{center}
  (e)
  \includegraphics[width=1.05\linewidth]{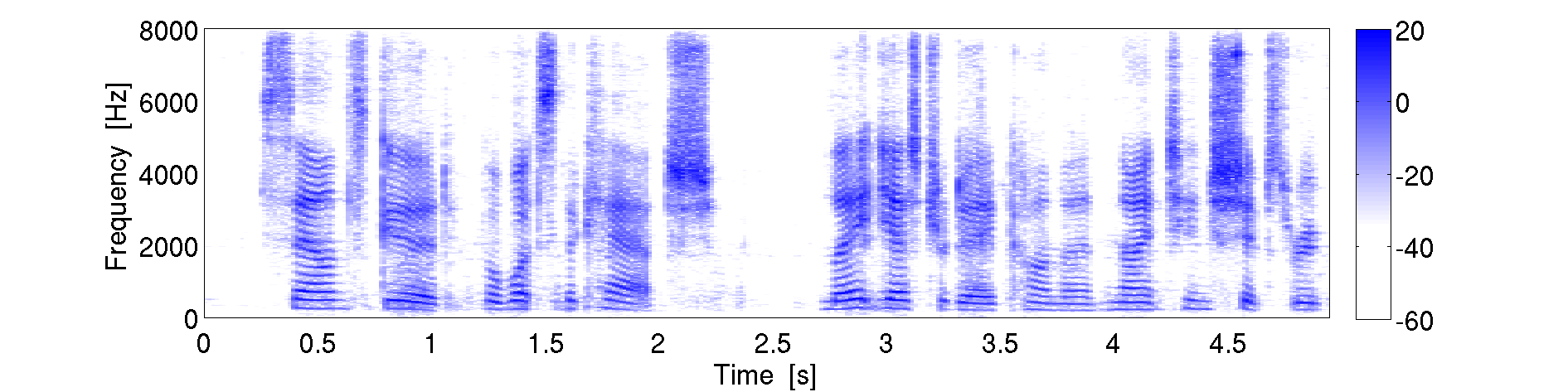}
  \end{center}
\end{minipage}

\caption{Performance plot of the BSSD-TD-SA model with $C=3$ speakers and \emph{real} RIRs. (a) STFT plot of the first microphone of the input mixture $\vm{z}(t)$. The $T_{60}$ of the reverb is approximately 650ms. (b) STFT plot of the clean first source signal $r_1(t)$. (c-e) STFT plots of the extracted and dereverberated sources $y_c(t)$.}
\label{fig:performance}
\end{figure}

Figure \ref{fig:performance} shows the performance of the BSSD-TD-SA model with the same three speakers as in Figure \ref{fig:doa_sphere}. From panel (a) it can be seen that there is a significant amount of reverberation in the input mixture $\vm{z}(t)$. Panel (b) and (d) show the extracted and dereverberated signals of two male speakers. Panel (c) shows the extracted and dereverberated signal of a female speaker. Even though speaker 2 and 3 are located very close to each other, they can still be separated. Note that the spatial resolution is only determined by the number of DOA vectors $D$, and the BSSD model is trained independently of the localization stage in Algorithm \ref{alg:localization}.


\subsection{Speaker Identification}

\begin{figure}[!ht]
\centering
\includegraphics[width=0.9\linewidth]{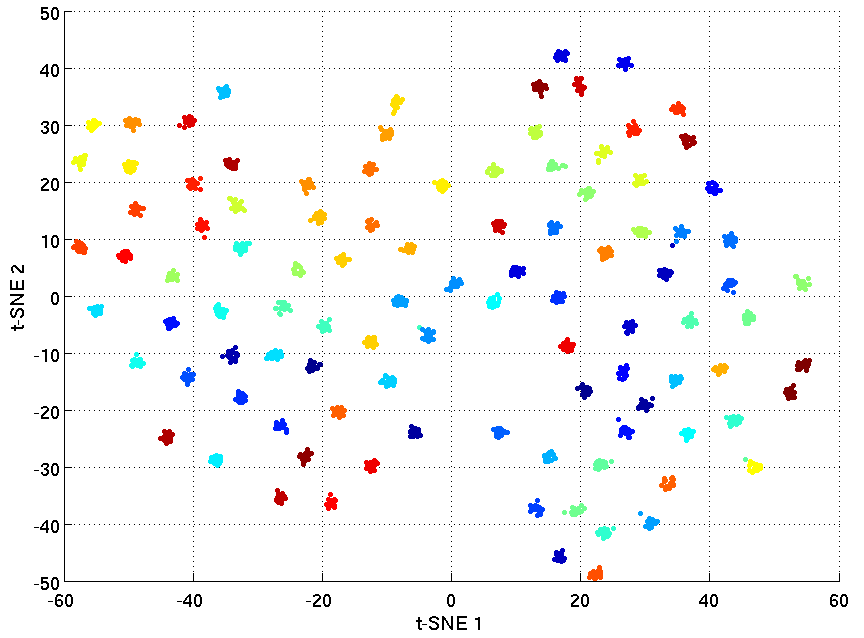}
\caption{t-SNE plot of the extracted speaker embeddings, using the 101 identities of the WSJ0 corpus.}
\label{fig:identification}
\end{figure}

Figure \ref{fig:identification} illustrates a 2D projection of the embeddings for each of the 101 WSJ0 speaker identities, which were obtained using BSSD-TD-SA model. For each speaker, 20 random utterances have been used, which have been reverberated and spatialized using both the \emph{real} and \emph{simulated} RIRs. It can be seen that each speaker identity can be distinguished.

\subsection{Offline mode}

Table \ref{tab:beamforming_real} reports SI-SDR, WER and EER for the \emph{real} RIRs. For $C=1$ speaker, the BSSD models only perform dereverberation. Hence, the SI-SDR is the highest for this case. The low WER of $9.65\%$ indicates that Google-ASR recognizes reverberant audio quite well. However, all BSSD models could lower the WER even further. Also, the EER is lowest for one speaker. This is to be expected, as no interfering components of other speakers reduce the quality of the speaker embeddings $\vm{e}$. For $C=2$ speakers, it can be seen that all BSSD models outperform Conv-TasNet and spatial PIT, even though Conv-TasNet is aided by WPE, and spatial PIT has explicitly been trained to perform dereverberation as well. Conv-TasNet could only achieve a WER of $48.71\%$. Spatial PIT only achieves a WER of $42.27\%$. It uses a static beamformer, which performs poorly in reverberant environments \cite{Pfei:aug2019}. 

For $C=3$ and $4$ speakers, performance of the BSSD architecture drops, as expected. I.e.: the SI-SDR gets lower, and both the WER and EER rise. The statistic adaption (SA) outperforms the analytic adaption (AA) variants for all number of speakers. This is also expected, as the SA variant allows the network to find an optimal transformation to separate the speakers both in time- and frequency domain, while the AA enforces a fixed scheme for spatial whitening and source localization. When comparing the time-domain (TD) to the frequency-domain (FD) variants, it can be seen that the FD models perform slightly better in terms of EER. This indicates that the speaker embeddings are easier to estimate in frequency domain, as they are calculated from the enhanced spectrograms (see Figure \ref{fig:bssd_fd}).

\begin{table}[!htb]
\begin{center}
  \caption{Speech separation, dereverberation and speaker identification performance for the \emph{real} RIRs in \emph{offline} mode.}
  
\setlength{\tabcolsep}{4pt}

\begin{tabular}{|l|c|r|r|r|}

\hline
\textbf{model} & $C$ & \textbf{SI-SDR} & \textbf{WER} & \textbf{EER} \\
\hline
\hline
no enhancement & 1 &  - & 9.65 \% &  - \\
\hline
Conv-TasNet & 2 & 6.92 dB & 48.71 \% &  - \\
\hline
spatial PIT & 2 & 2.26 dB & 42.27 \% &  - \\
\hline
\multirow{4}{*}{BSSD-FD-AA} & 1 & 16.91 dB &  5.09 \% & 2.87 \% \\
                            & 2 &  8.65 dB & 28.74 \% & 5.92 \% \\
                            & 3 &  6.75 dB & 51.90 \% & 8.94 \% \\
                            & 4 &  5.61 dB & 66.20 \% & 11.32 \% \\
\hline
\multirow{4}{*}{BSSD-FD-SA} & 1 & 14.92 dB &  6.10 \% & 3.02 \% \\
                            & 2 & 10.23 dB & 23.70 \% & 4.22 \% \\
                            & 3 &  8.34 dB & 42.49 \% & 6.20 \% \\
                            & 4 &  7.17 dB & 56.43 \% & 7.22 \% \\
\hline
\multirow{4}{*}{BSSD-TD-AA} & 1 & 10.07 dB &  9.81 \% & 4.18 \% \\
                            & 2 &  6.74 dB & 45.79 \% & 9.94 \% \\
                            & 3 &  5.22 dB & 71.63 \% & 15.68 \% \\
                            & 4 &  4.31 dB & 84.39 \% & 21.65 \% \\
\hline
\multirow{4}{*}{BSSD-TD-SA} & 1 & 14.40 dB &  5.72 \% & 2.89 \% \\
                            & 2 &  9.33 dB & 26.19 \% & 5.75 \% \\
                            & 3 &  7.92 dB & 42.32 \% & 7.28 \% \\
                            & 4 &  6.84 dB & 56.57 \% & 9.39 \% \\
\hline

\end{tabular}

  \label{tab:beamforming_real}
\end{center}
\end{table}

Table \ref{tab:beamforming_simu} reports SI-SDR, WER and EER for the significantly shorter \emph{simulated} RIRs. Consequently, all systems perform better in all scores. In these almost ideal conditions, Google-ASR achieved a WER of $3.04\%$ for a single speaker without any enhancement. However, all BSSD variants also achieved a  lower WER as well. Also, Conv-TasNet and spatial PIT perform better compared to the \emph{real} RIRs. However, Conv-TasNet still could not separate the speakers perfectly. The static beamformer of spatial PIT performs quite well for the short \emph{simulated} RIRs, achieving a WER of $17.1\%$. However, all BSSD variants achieved a lower WER for $C=2$ speakers. Again, the FD variants perform slightly better than the TD models, and the SA layer outperforms the AA layer.

\begin{table}[!htb]
\begin{center}
  \caption{Speech separation, dereverberation and speaker identification performance for the \emph{simulated} RIRs in \emph{offline} mode.}
  
\setlength{\tabcolsep}{4pt}

\begin{tabular}{|l|c|r|r|r|}

\hline
\textbf{model} & $C$ & \textbf{SI-SDR} & \textbf{WER} & \textbf{EER} \\
\hline
\hline
no enhancement & 1 &  - & 3.04 \% &  - \\
\hline
Conv-TasNet & 2 & 8.78 dB & 25.34 \% &  - \\
\hline
spatial PIT & 2 & 3.06 dB & 17.10 \% &  - \\
\hline
\multirow{4}{*}{BSSD-FD-AA} & 1 & 22.72 dB &  1.72 \% & 2.89 \% \\
                            & 2 & 10.93 dB & 14.71 \% & 6.11 \% \\
                            & 3 &  8.62 dB & 27.82 \% & 8.27 \% \\
                            & 4 &  7.25 dB & 37.58 \% & 9.65 \% \\
\hline
\multirow{4}{*}{BSSD-FD-SA} & 1 & 22.02 dB &  2.72 \% & 3.07 \% \\
                            & 2 & 12.06 dB & 12.80 \% & 5.25 \% \\
                            & 3 &  9.06 dB & 25.39 \% & 7.37 \% \\
                            & 4 &  7.40 dB & 40.36 \% & 8.99 \% \\
\hline
\multirow{4}{*}{BSSD-TD-AA} & 1 & 16.87 dB &  2.71 \% & 3.44 \% \\
                            & 2 & 10.62 dB & 15.55 \% & 7.23 \% \\
                            & 3 &  8.31 dB & 30.86 \% & 10.05 \% \\
                            & 4 &  6.75 dB & 46.48 \% & 14.52 \% \\
\hline
\multirow{4}{*}{BSSD-TD-SA} & 1 & 22.75 dB &  2.02 \% & 2.84 \% \\
                            & 2 & 12.82 dB &  9.84 \% & 5.56 \% \\
                            & 3 & 10.23 dB & 20.92 \% & 7.77 \% \\
                            & 4 &  8.57 dB & 34.45 \% & 9.23 \% \\
\hline

\end{tabular}

  \label{tab:beamforming_simu}
\end{center}
\end{table}

\subsection{Block-online mode}

Table \ref{tab:localization_real} reports SI-SDR, WER and BER for the \emph{real} RIRs, for block lengths of $T_B = 1s$, $2.5s$ and $5s$. We only performed these experiments on the SA variants of the BSSD network, as the SA layer consistently outperforms the AA layer. The SI-SDR and WER are worse compared to \emph{offline} mode, as many sources of errors cumulate throughout the processing chain. I.e.: Algorithm \ref{alg:localization} may produce wrong DOA indices for short blocks and many speakers. Consequently, speaker separation is poor, resulting in erroneous speaker embeddings. Further, both the speaker separation and speaker identification modules introduce errors on their own. For the shortest block length of $T_B=1s$, there are 20 blocks for $20s$ of audio. In order to achieve a perfect BER score, the embeddings for the same speaker in all 20 blocks must be identical (see Eq. (\ref{eq:ber})). If the speaker is silent in one or more blocks, a perfect BER score cannot be achieved. Clearly, performance is better for larger block lengths and fewer speakers. For $C=2$ speakers and a block length of $T_B=5s$, the WER is $34.79\%$ for the FD variant, and $28.51\%$ for the TD variant. The BER is $10\%$ for the FD variant, and $4.5\%$ for the TD variant. In contrast to the experiments in \emph{offline} mode, all scores are slightly better for the TD models.

\begin{table}[!htb]
\begin{center}
  \caption{Speech separation and dereverberation performance for the \emph{real} RIRs in \emph{block-online} mode.}
  
\setlength{\tabcolsep}{4pt}

\begin{tabular}{|l|c|r|r|r|r|}

\hline
\textbf{model} & $C$ & $T_B$ & \textbf{SI-SDR} & \textbf{WER} & \textbf{BER} \\
\hline
\hline
\multirow{9}{*}{BSSD-FD-SA} & \multirow{3}{*}{2} & 1.0s &  3.80 dB & 66.94 \% & 37.40 \% \\
                            &                    & 2.5s &  7.05 dB & 44.22 \% & 15.88 \% \\
                            &                    & 5.0s &  8.64 dB & 34.79 \% & 10.00 \% \\
\cline{2-6}
                            & \multirow{3}{*}{3} & 1.0s &  3.00 dB & 75.68 \% & 48.90 \% \\
                            &                    & 2.5s &  5.19 dB & 59.94 \% & 27.13 \% \\
                            &                    & 5.0s &  6.73 dB & 52.98 \% & 14.75 \% \\
\cline{2-6}
                            & \multirow{3}{*}{4} & 1.0s &  2.49 dB & 78.21 \% & 64.40 \% \\
                            &                    & 2.5s &  3.71 dB & 71.96 \% & 43.75 \% \\
                            &                    & 5.0s &  4.76 dB & 68.07 \% & 35.00 \% \\
\hline
\multirow{9}{*}{BSSD-TD-SA} & \multirow{3}{*}{2} & 1.0s &  4.49 dB & 60.67 \% & 25.30 \% \\
                            &                    & 2.5s &  7.04 dB & 36.24 \% & 10.75 \% \\
                            &                    & 5.0s &  8.47 dB & 28.51 \% & 4.50 \% \\
\cline{2-6}
                            & \multirow{3}{*}{3} & 1.0s &  3.30 dB & 74.11 \% & 39.70 \% \\
                            &                    & 2.5s &  4.81 dB & 63.82 \% & 23.88 \% \\
                            &                    & 5.0s &  6.31 dB & 48.68 \% & 22.50 \% \\
\cline{2-6}
                            & \multirow{3}{*}{4} & 1.0s &  2.70 dB & 76.32 \% & 47.85 \% \\
                            &                    & 2.5s &  3.93 dB & 70.83 \% & 32.25 \% \\
                            &                    & 5.0s &  4.86 dB & 65.14 \% & 32.50 \% \\
\hline

\end{tabular}

  \label{tab:localization_real}
\end{center}
\end{table}

Table \ref{tab:localization_simu} reports SI-SDR, WER and BER for the \emph{simulated} RIRs, for block lengths of $T_B = 1s$, $2.5s$ and $5s$. All scores are better compared to the significantly longer \emph{real} RIRs. For $C=2$ speakers and a block length of $T_B=5s$, the WER is $19.80\%$ for the FD variant, and $16.75\%$ for the TD variant. The BER is $3.25\%$ for the FD variant, and $1.25\%$ for the TD variant. Again, In contrast to the experiments in \emph{offline} mode, all scores are slightly better for the TD models.

\begin{table}[!htb]
\begin{center}
  \caption{Speech separation and dereverberation performance for the \emph{simulated} RIRs in \emph{block-online} mode.}
  
\setlength{\tabcolsep}{4pt}

\begin{tabular}{|l|c|r|r|r|r|}

\hline
\textbf{model} & $C$ & $T_B$ & \textbf{SI-SDR} & \textbf{WER} & \textbf{BER} \\
\hline
\hline
\multirow{9}{*}{BSSD-FD-SA} & \multirow{3}{*}{2} & 1.0s &  4.24 dB & 58.08 \% & 27.55 \% \\
                            &                    & 2.5s &  8.13 dB & 27.70 \% & 7.88 \% \\
                            &                    & 5.0s & 10.67 dB & 19.80 \% & 3.25 \% \\
\cline{2-6}
                            & \multirow{3}{*}{3} & 1.0s &  3.49 dB & 75.03 \% & 40.00 \% \\
                            &                    & 2.5s &  6.46 dB & 46.98 \% & 17.50 \% \\
                            &                    & 5.0s &  7.91 dB & 35.83 \% & 11.75 \% \\
\cline{2-6}
                            & \multirow{3}{*}{4} & 1.0s &  2.89 dB & 81.74 \% & 49.40 \% \\
                            &                    & 2.5s &  5.25 dB & 52.42 \% & 24.63 \% \\
                            &                    & 5.0s &  6.05 dB & 49.52 \% & 23.75 \% \\
\hline
\multirow{9}{*}{BSSD-TD-SA} & \multirow{3}{*}{2} & 1.0s &  5.82 dB & 51.17 \% & 18.55 \% \\
                            &                    & 2.5s & 10.94 dB & 18.21 \% & 3.63 \% \\
                            &                    & 5.0s & 11.91 dB & 16.75 \% & 1.25 \% \\
\cline{2-6}
                            & \multirow{3}{*}{3} & 1.0s &  4.40 dB & 73.55 \% & 30.50 \% \\
                            &                    & 2.5s &  7.75 dB & 47.37 \% & 15.25 \% \\
                            &                    & 5.0s &  9.66 dB & 39.12 \% & 9.75 \% \\
\cline{2-6}
                            & \multirow{3}{*}{4} & 1.0s &  3.10 dB & 81.24 \% & 37.65 \% \\
                            &                    & 2.5s &  5.11 dB & 60.12 \% & 26.00 \% \\
                            &                    & 5.0s &  7.80 dB & 52.99 \% & 10.75 \% \\
\hline

\end{tabular}

  \label{tab:localization_simu}
\end{center}
\end{table}

\subsection{Model complexity}

Table \ref{tab:model_complexity} reports the number of trainable parameters per variant of the BSSD network. The frequency domain (FD) variants use mostly complex-valued weights, which are counted as 2 real-valued weights. Hence, these models are significantly larger than the time domain (TD) variants. The size of the statistic adaption (SA) layer in the time domain network is comparatively small with 720,000 parameters. However, the analytic adaption (AA) layer requires additional convolutions from Eq. (\ref{eq:analytic_adaption_td}). Similar to Conv-TasNet, the time domain variant also has the advantage of a small step size of 50 samples. The number of parameters for speaker identification is almost the same for all variants.

\begin{table}[!htb]
\begin{center}
  \caption{Number of parameters for the \emph{beamforming} and \emph{identification} branches of the BSSD network.}
  
\newcolumntype{L}[1]{>{\raggedright\let\newline\\\arraybackslash\hspace{0pt}}m{#1}}
\newcolumntype{C}[1]{>{\centering\let\newline\\\arraybackslash\hspace{0pt}}m{#1}}
\newcolumntype{R}[1]{>{\raggedleft\let\newline\\\arraybackslash\hspace{0pt}}m{#1}}

\setlength{\tabcolsep}{4pt}

\begin{tabular}{|l|R{15mm}|R{20mm}|}

\hline
\textbf{model} & \textbf{parameters beamformer} & \textbf{parameters identification} \\
\hline
\hline
BSSD-FD-AA & 11,064,384 & 2,664,100 \\
BSSD-FD-SA & 14,757,984 & 2,664,100 \\
BSSD-TD-AA &  5,456,700 & 2,526,500 \\
BSSD-TD-SA &  6,176,700 & 2,526,500 \\
\hline

\end{tabular}

  \label{tab:model_complexity}
\end{center}
\end{table}

\section{Conclusion}
\label{sec:conclusion}

In this paper, we introduced the \emph{Blind Speech Separation and Dereverberation} (BSSD) network, which performs simultaneous \emph{speaker separation}, \emph{dereverberation} and \emph{speaker identification} in a single neural network. We proposed four variants of our system, which operate in frequency-domain and time-domain, and use analytic adaption and statistic adaption layers to perform blind speaker separation. We have shown that 100 DOA bases provide enough spatial resolution to separate up to four speakers. Further, we proposed the \emph{block-online} mode to process longer audio recordings, as they occur in meeting scenarios. In our experiments, we could show that the BSSD network outperforms similar state-of-the art algorithms for speaker separation in terms of SI-SDR and WER.

\bibliographystyle{IEEEtran}
\bibliography{bibliography}

\vspace{-5mm}
\begin{IEEEbiography}[{\includegraphics[width=2cm]{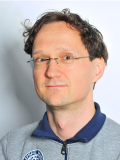}}]{Lukas Pfeifenberger} received the M.Sc. (Dipl. Ing.FH) degree in computer science from the University of Applied Sciences, Salzburg, Austria, in 2004. His master's thesis promotes the use of closed control systems in earth fault detection appliances. Since 2005 he has been working in the electronics industry on projects pertaining to FPGA design, DSP programming and communication acoustics, including algorithms for echo and noise cancellation. In 2013, he received the M.Sc. degree in Telematics at Graz University of Technology, Austria. His master's thesis decuments the implementation of an acoustic beamformer on an embedded device with limited resources. Since 2015 he has been a Research Associate at the Laboratory of Signal Processing and Speech Communication, Graz University of Technology, Austria. In 2021, he received his Ph.D. degree in computer science with honors from Graz University of Technology, Austria. His doctoral thesis explores the evolution of neural acoustic beamformers ranging from mask-based beamforming to blind speaker separation. His research interests include signal processing, machine learning, artificial intelligence, pattern recognition, computer vision, speech enhancement, acoustic echo control, speaker separation, and data analysis for industrial applications. 
\end{IEEEbiography}

\vspace{-10mm}
\begin{IEEEbiography}[{\includegraphics[width=2cm]{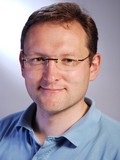}}]{Franz Pernkopf}
received his MSc (Dipl. Ing.) degree in Electrical Engineering at Graz University of Technology, Austria, in summer 1999. He earned a PhD degree from the University of Leoben, Austria, in 2002. In 2002 he was awarded the Erwin Schr\"{o}dinger Fellowship. He was a Research Associate in the Department of Electrical Engineering at the University of Washington, Seattle, from 2004 to 2006. From 2010-2019 he was Associate Professor at the Laboratory of Signal Processing and Speech Communication, Graz University of Technology, Austria. Since 2019, he is Professor for Intelligent Systems at the Signal Processing and Speech Communication Laboratory at Graz University of Technology, Austria. His research is focused on pattern recognition, machine learning, and computational data analytics with applications in signal and speech processing.
\end{IEEEbiography}


\end{document}